\documentclass[12pt]{article}
\usepackage{amsmath}
\usepackage{amscd}
\usepackage{array}
\usepackage{latexsym,amssymb}
\newcommand{\nn}{\nonumber}

\def\sx{\left}
\def\dx{\right}
\def\to{\rightarrow}

\def\cT{\mathcal{T}}

\newcommand{\eq}[1]{(\ref{#1})}

\newcommand{\be}{\begin{equation}}
\newcommand{\ee}{\end{equation}}
\newcommand{\bea}{\begin{eqnarray}}
\newcommand{\eea}{\end{eqnarray}}

\newcommand{\ba}{\begin{eqnarray}}
\newcommand{\ea}{\end{eqnarray}}

\def\a{\alpha}

\def\b{\beta}

\def\d{\delta}
\def\e{\epsilon}

\def\g{\gamma}

\def\l{\lambda}

\def\o{\omega}

\def\s{\sigma}

\def\D{\Delta}
\def\G{\Gamma}
\def\L{\Lambda}

\def\S{\Sigma}

\def\cD{\mathcal{D}}

\def\cF{\mathcal{F}}
\def\cG{\mathcal{G}}

\def\cL{\mathcal{L}}
\def\cM{\mathcal{M}}
\def\cN{\mathcal{N}}

\def\cQ{\mathcal{Q}}
\def\cR{\mathcal{R}}
\def\cS{\mathcal{S}}

\def\re{{\rm Re}\mathcal{N}}

\def\ii{\mathrm{i}}
\def\ib{{\ol {\imath}}}
\def\j{\jmath}
\def\jb{{\ol {\jmath}}}
\def\kb{{\ol  k}}
\def\lb{{\ol  \ell}}
\def\mb{{\ol  m}}
\def\nb{{\ol {n}}}
\def\rb{{\ol {r}}}

\def\notin{\hbox{{$\in$}\kern-.51em\hbox{/}}}

\def\inbar{\vrule height1.5ex width.4pt depth0pt}
\def\IB{\relax{\rm I\kern-.18em B}}
\def\IC{\relax\,\hbox{$\inbar\kern-.3em{\rm C}$}}
\def\ID{\relax{\rm I\kern-.18em D}}
\def\IE{\relax{\rm I\kern-.18em E}}
\def\IF{\relax{\rm I\kern-.18em F}}
\def\IG{\relax\,\hbox{$\inbar\kern-.3em{\rm G}$}}
\def\IH{\relax{\rm I\kern-.18em H}}
\def\II{\relax{\rm I\kern-.17em I}}
\def\IK{\relax{\rm I\kern-.18em K}}
\def\IL{\relax{\rm I\kern-.18em L}}
\def\IN{\relax{\rm I\kern-.18em N}}
\def\IP{\relax{\rm I\kern-.18em P}}
\def\IQ{\relax\,\hbox{$\inbar\kern-.3em{\rm Q}$}}
\def\IR{\relax{\rm I\kern-.18em R}}
\def\IU{\relax\,\hbox{$\inbar\kern-.3em{\rm U}$}}
\def\ZZ{\relax\ifmmode\mathchoice{\hbox{\cmss
Z\kern-.4em Z}}{\hbox{\cmss Z\kern-.4em Z}}{\lower.9pt\hbox{\cmsss Z\kern-.4em Z}} {\lower1.2pt\hbox{\cmsss
Z\kern-.4em Z}}\else{\cmss Z\kern-.4em Z}\fi}
\def\IGam{\relax{{\rm I}\kern-.18em \Gamma}}

\newcommand{\iden}{\mbox{{1}\hspace{-.11cm}{l}}}
\def\bfnull{\relax{\rm O \kern-.635em 0}}

\def\de{{\rm d}}%\hskip -1pt}
\def\der{\partial}

\def\na{\nabla}

\def\ol{\overline}

\def\imez{\frac{{\rm i}}{2}}
\def\mez{\frac{1}{2}}
\def\qu{\frac{1}{4}}

\def\square{{\,\lower0.9pt\vbox{\hrule
\hbox{\vrule height 0.2 cm \hskip 0.2 cm \vrule height 0.2 cm}\hrule}\,}}

\def\twomat#1#2#3#4{\left(\begin{array}{cc} \end{array} \right)}

\begin{document}

\numberwithin{equation}{section}

\begin{center}
{\Large
$D=4$, $N=2$ Gauged Supergravity coupled to\\
Vector-Tensor   Multiplets }

\vskip 1.5cm

{\bf \large Laura Andrianopoli$^{1}$, Riccardo D'Auria$^1$,\\
  Luca Sommovigo$^2$ and Mario Trigiante$^{1}$}\\
\vskip 8mm
 \end{center}
\noindent {\small{\it $^1$  Dipartimento di Fisica, Politecnico di Torino, Corso Duca
    degli Abruzzi 24, I-10129 Turin, Italy and Istituto Nazionale di
    Fisica Nucleare (INFN) Sezione di Torino, Italy; E-mail:  {\tt
      laura.andrianopoli@polito.it}; {\tt riccardo.dauria@polito.it}; {\tt mario.trigiante@polito.it}\\
    $^2$ DISTA, Universit\`a del Piemonte Orientale, Viale Teresa Michel 11, 15121 Alessandria, Italy; E-mail: {\tt luca.sommovigo@mfn.unipmn.it}}}

\vfill

\begin{center}
{\bf Abstract}
\end{center}
{\small We construct the general four-dimensional $N=2$ supergravity theory coupled to vector and vector-tensor
multiplets only. Consistency of the construction requires the introduction of the vector fields dual to those
sitting in the same supermultiplets as the antisymmetric tensors, as well as the scalar fields dual to the
tensors themselves. Gauge symmetries also involving these additional fields guarantee the correct counting of
the physical degrees of freedom. }

\vfill\eject

%%%%%%%%%%%%%%%%%%%%%%%%%%%%%%%%%%%%%%%%%%%%%%%%%%%%%
%%%%%%%%%%%%%%%%%%%%%%%%%%%%%%%%%%%%%%%%%%%%%%%%%%%%%
%%%%%%%%%%%%%%%%%%%%%%%%%%%%%%%%%%%%%%%%%%%%%%%%%%%%%

%\begin{center}
%\today
%\end{center}

%%%%%%%%%%%%%%%%%%%%%%%%%%%%%%%%%%%%%%%%%%%%%%%%%%%%

\section{Introduction}
\label{intro}

In the last decade, a considerable interest has been devoted to investigating  the role of antisymmetric tensors
($p$-forms with $p>1$) in four and five dimensional extended supergravity theories
\cite{dewit,Gunaydin:1999zx,lm,vandoren,Dall'Agata:2003yr,D'Auria:2004yi,Sommovigo:2004vj,Sommovigo:2005fk,
D'Auria:2004sy,Bergshoeff:2004kh,deWit:2004nw,deWit:2005ub,de
Wit:2007mt,mip,D'Auria:2007ay,de Vroome:2007zd,deWit:2008ta,ortin}. The study of Free Differential Algebras including
gauge-coupled forms of higher order has an interest per se: It has been shown that for these theories the
consistency of the 1-form sector does not imply  the Jacobi identities, which indeed are no longer satisfied
\cite{de Wit:2007mt}. However  it is still possible, via a  field redefinition, to choose a setting where the
Jacobi identities are satisfied, at the price of having a deformation of the gauge couplings in the Bianchi
identities \cite{mip}.

Moreover, antisymmetric tensors naturally appear in string compactifications and in supergravity theories with
fluxes turned on. Focussing on four dimensional N-extended theories, the relevant tensors are given by 2-forms
which, at the ungauged level, can always be dualized to real scalars. However, when gauge charges and masses are
turned on, theories with tensor multiplets contain different couplings with respect to the ones where all
tensors are dualized. In particular, the masses of the tensors appear as magnetic charges, so that the
symplectic frame involved is in general quite different with respect to the standard theories. Some years ago,
the N=2 theory coupled to hyper-tensor multiplets (obtainable from hypermultiplets upon dualizing some of the
scalar fields into antisymmetric tensors \cite{Sommovigo:2005fk}) was constructed
\cite{vandoren,Dall'Agata:2003yr,D'Auria:2004yi}. It exhibited a symplectically invariant scalar potential  in the sector
related to the $SU(2)$ part of the R-symmetry. Eventually, the N=2 theory in five dimensions coupled to hyper-,
vector-, and massive hypertensor multiplets was constructed \cite{Gunaydin:1999zx,Bergshoeff:2004kh} and a
symplectically covariant formulation of maximally extended theories in five and four dimensions, involving
tensor fields, was given in \cite{deWit:2004nw,deWit:2005ub,de Wit:2007mt}, making use of the embedding tensor
formalism \cite{Cordaro:1998tx,Nicolai:2000sc,deWit:2002vt}. Few years ago,  some of the present authors studied
on general grounds the Free Differential Algebra of forms of various degree based on the gauging of a general
Lie algebra $G$ \cite{mip} focussing on the Higgs mechanism through which the higher order forms get their mass.
In this work, just as in \cite{deWit:2004nw},  the ``selfduality in odd dimensions'' mechanism
\cite{Townsend:1983xs}, on which \cite{Bergshoeff:2004kh} was based,  originated from the gauge-fixing  of a
theory formulated in terms of gauge-coupled massless fields, via the Higgs mechanism.  For the $N=2$  theory in
five dimensions, it was also pointed out in \cite{mip} that the Higgs mechanism has to be at work all
\emph{within} the same multiplet, so that the massive tensor multiplets have to be BPS, i.e. short ones. In
particular, the 2-form $B_M$ acquires mass by ``eating'' the degrees of freedom of its Hodge-dual vector $A^M$.

 A complete general formulation of N=2 supergravity
 in four dimensions coupled to vector-tensor multiplets was still missing.
 Some work in that direction was  done in \cite{dewit}, where the coupling of $N=2$ supergravity to a vector-tensor multiplet was first studied.
Afterwards,  in \cite{Gunaydin:2005df,Gunaydin:2005bf}  direct compactification
 of the five dimensional Lagrangian was given, and in \cite{mip} the supersymmetric Bianchi identities
  were solved up to three-fermions terms, giving the expression of the scalar potential  and a set of constraints
  on the geometry of the $\sigma$-model.

 The aim of this paper is to write the general Lagrangian of  N=2, D=4 supergravity coupled to vector multiplets
 and vector-tensor multiplets. We find that, as for the five dimensional case,  the tensor  has to belong to a short
 representation of supersymmetry \footnote{Note that this implies the presence of a central charge, in agreement with the results in \cite{dewit}.}.
  In this case, as will be discussed in the text, the Hodge duality acts non trivially: the tensor, $B_M$,
   becomes massive  by eating the degrees of freedom of a vector, $A^M$, which is the Hodge-dual of the vector $A_M$
   in the multiplet; the vector $A_M$ in turn gets mass by eating the degree of freedom of the scalar Hodge-dual to
   the tensor $B_M$. This in turn implies that the symplectic embedding is quite involved.
   A similar mechanism was described  in \cite{deWit:2005ub,de Wit:2007mt}.
  For this reason, the construction of the model requires to work in an enlarged field space where the Hodge-dual
  fields involved in the gauging are present together with the fields composing the multiplets. As a consequence,
  requiring supersymmetry and gauge invariance, we obtain a set of constraints on the fields and in particular the ones
  determining the geometry of the $\sigma$-model, which still need to be explicitly solved.
  The explicit analysis of the geometry of the manifold spanned by the scalars which
survive the dualization into tensor fields is postponed   to a forthcoming publication \cite{forth}. Moreover,
the Lagrangian we get has a manifest symplectic invariance in the sector involving vector-tensor multiplets. It
would be interesting to extend the symplectic invariance also to the vector multiplets sector. This would
require, as explained in \cite{Sommovigo:2004vj,Dall'Agata:2003yr,D'Auria:2004yi},  the coupling of the theory
also  to hypermutiplets and hypertensor multiplets in a non abelian way. Alternatively, it could be found as a
symplectically covariant gauging of the standard general matter coupled $N=2$ supergravity, in the spirit of the
approach followed in \cite{de Wit:2007mt} for the $N=8$ theory. The extension of the construction to include
hypermultplets and FI terms, is left to future investigation.

The paper is organized as follows: \\
Section \ref{susy4} contains a description of the field content and of the peculiarities of the theory. \\
In section \ref{bosfda} we discuss  the bosonic Free Differential Algebra underlying our theory, studying in particular the gauge structure and the symplectic embedding.\\
Sections  \ref{theory} and \ref{comments} include  the main results of our paper: \\
 Section \ref{theory} contains the complete Lagrangian and supersymmetry transformation laws, while in Section  \ref{comments} we comment on the results found and make some observations on the geometry of the embedding scalar manifold and on its relation with Special Geometry. \\
The Appendices contain technical and computational details:\\
 Appendix \ref{algebraemb} contains an explicit discussion of closure of the gauge algebra when embedded in the symplectic algebra.\\
In Appendix \ref{susyparam} we list the superspace Bianchi identities of the fields together with their rheonomic parametrizations.\\
In Appendix \ref{Lagrangian} we give the superspace rheonomic Lagrangian of the theory.\\
In Appendix \ref{relations} we collect all the constraints found on the fields, on the $\sigma$-model and on the gauging from solving the superspace Bianchi identities and the superspace field equations from the rheonomic Lagrangian.\\
Finally in Appendix \ref{dualizationsection} we study in detail the dualization procedure, showing that our kinetic Lagrangian can be obtained from that of standard $N=2$ supergravity by dualization of some of the scalars in the vector multiplets into 2-form tensors.

%%%%%%%%%%%%%%%%%%%%%%%%%%%%%%%%%%%%%%%%%%%%%%%%%%%%%%%%%%%

\section{The vector-tensor multiplet structure} \label{susy4}
 Let us consider $N=2$ supergravity in four dimensions with field content given by:
 \begin{itemize}
 \item The gravity multiplet:
$$(V^a_\mu, \psi_{A\mu} ,\psi_\mu^A , A^0_\mu)\,, $$ (where $a$ and $\mu$ denote space-time indices respectively
flat and curved, $A=1,2$ is an R-symmetry $SU(2)$ index and we have decomposed the gravitino in chiral
($\psi_A$) and anti-chiral ($\psi^A$) components),
\item $n_V$ vector multiplets: $$(A_\mu, \lambda^{A},
\lambda_A, z)^r\,,\qquad r=1,\cdots,n_V\,,$$ where $z^r$ are holomorphic coordinates on the special manifold
$\cM_V$ spanned by its scalar sector and $\lambda^{rA}$ are chiral spin-1/2 fields (with complex conjugate
antichiral component $\lambda^\rb_A$),
\item  $n_T$ vector-tensor multiplets: $$\left(B_{\mu\nu},A_{\mu},
\chi^A, \chi_{ A}, P \right)_M\,,\qquad M=1,\cdots,n_T\,,$$ where $P_M$ are real functions of the scalar
 fields spanning the real manifold $\cM_T$, which can be chosen as ``special'' coordinates on $\cM_T$. Here
 $M$ is a representation index of the gauge group $G$.   $\chi_M^A,\chi_{M A}$ denote left- and right-hand
   components of Majorana spinors respectively.
   \end{itemize}

This theory can be thought of as obtained from standard $N=2$ supergravity coupled to vector multiplets by
Hodge-dualization of, say, the imaginary part of a subset of the complex scalar fields parametrizing the special
manifold. More precisely, starting from $n=n_V+n_T$ vector multiplets of $N=2$ supergravity with scalar fields
$z^i= (z^r,z^m)$ (with $m=1,\cdots ,n_T$, $i=1,\cdots ,n_V+n_T$), we shall find that the Hodge duality between
tensor and scalar fields is given by:
\begin{equation}
H_{M|\mu\nu\rho}= -\frac \ii 3\sqrt{-g}\,\epsilon_{\mu\nu\rho\sigma}(p_{Mi}\nabla^\sigma z^i- \bar p_{M\ib}\nabla^\sigma \bar z^\ib)
\label{dualization}\end{equation}
 $p_{Mi}(z,\bar z)= \nabla_i P_M$ converting coordinate indices into representation indices
  of the gauge group of the theory.  This is an on-shell equation obtained from closure of Bianchi identities
   in superspace or, equivalently, from the equations of motion of the rheonomic Lagrangian. From the space-time
   point of view, this is equivalent to requiring closure of the supersymmetry algebra.

Note that \eq{dualization} does not identify the Hodge dual of $H_M$ with an exact form, say $dY_M$ but, for the
sake of simplicity, we shall find it convenient to refer to the degrees of freedom dual to the tensor fields as
$Y_M^{(1)}=Y_{Mm}dy^m$.

To our knowledge, as anticipated in the introduction, the construction of such theory in full generality has not
been achieved so far, even if important steps in that direction have been given in ref. \cite{Gunaydin:2005bf},
where the four dimensional theory was obtained by dimensional reduction from five dimensions and the ensuing
properties thoroughly analyzed. However, that approach does not catch the most general theory,  being restricted
to models with a five dimensional uplift. Finally, in \cite{mip} a relevant part of the construction has been carried
out by some of the authors. In particular, in that paper we discussed  the solution of Bianchi identities in
superspace which, besides giving the general supersymmetry transformation laws and the constraints on the
geometry of the relevant $\sigma$-models, also allows us to retrieve in principle the equations of motion of the
theory.

In \cite{mip} we solved, up to three fermions, the Free Differential Algebra Bianchi identities in superspace
following the so-called geometric (or rheonomic) approach of \cite{D'Auria:1980ra}. This will be the starting
point of our development here. Since  some of the notations and conventions have been changed here with respect
to \cite{mip}, for the benefit of the reader we  expose in the present paper the main results found there. We
recall that, in order for the Free Differential Algebra to close in superspace, it is necessary to include among
the defining bosonic fields of the tensor multiplet sector, besides the vectors $A_M$ and the tensors $B_M$,
also their Hodge duals, that is the (auxiliary) vectors $A^M$, Hodge dual to the $A_M$, and the real scalars
$y^m$, Hodge dual to the tensors $B_M$. The gauge group $G$ is gauged by the  vectors $A^\Lambda=(A^X,A^M)$,
that is by the $A^X\equiv (A^0,A^r$), $A^0$ being
  the graviphoton,
  together with the $A^M$.

   Let us remark that, if we think of the theory as dualization of ordinary $N=2$ supergravity coupled to
   vector multiplets,
     the theory with tensor
   multiplets is in a rotated symplectic frame. Depending on whether the theory is thought of as constructed directly
   from vector-tensor multiplets or as dualization of standard $N=2$, the interpretation of the
   vectors $A^M,A_M$ is different. Indeed, in the former case the $A_M$ are the physical fields, to be
   considered as electric, while the gauge group includes the magnetic vectors $A^M$. On the other hand, in the
   second interpretation the $A^M$ are electric and the $A_M$ are magnetic fields.
   %could be obtained  from an "electric" theory including only vector multiplets ($n_V+n_T$ vector multiplets)
%   by Hodge-dualization, in $n_T$ multiplets, of one of the two real scalars in $z^m$, $Y^M$, into the 2-index tensor $B_M$ and of the "electric" vector $A^M$ into the "magnetic" one $A_M$.
In writing the Lagrangian, we will consider  $A^M$ as the propagating gauge fields. It will  be useful to adopt
a collective gauge-vector index $\L=(X,M)=0,1,\cdots,n_V +n_T$ (with $X =0,1,\cdots,n_V$) running over  the
corresponding vectors of the theory. In the study of  the supersymmetric Free Differential Algebra of the
theory, we shall let all the vectors $A^{\L}$ be the gauge vectors of a non abelian algebra $G$ and the tensors
$B_M$   be in a representation of it \footnote{As discussed in \cite{mip}, starting from a general algebra $G$,
with the constraint on the structure constants $f_{\Lambda M}{}^X=0$, we can always retrieve, by a suitable
redefinition of the 2-forms $B_M$, the case $G=G_0\ltimes \IR^{n_T}$, where $G_0$ is in general   a contraction
of $G$ gauged by the vectors $A^X$.}.

 In the interacting theory, the Higgs mechanism   takes place so that the
vectors $A^M$   provide the degrees of freedom giving mass to the tensors $B_M$. In this way the gauge algebra
is broken to a particular contraction $G_0$ ($\dim G_0= n_V+1$) spanned  by the vectors $(A^0, A^r)$.
 On the other hand, the  gauge vectors $A_M$ undergo a dual Higgs
mechanism, since they take mass by eating the degrees of freedom of the dualized scalars $y^m$, and they  will
appear in the supercurvature of the tensor field-strengths $H_M$. As already remarked, if we did not include all
the fields together with their Hodge duals,  inconsistencies would show up in the superspace Bianchi identities.
Note that our approach has been to introduce the dual fields as auxiliary fields, letting  closure of the free
differential algebra and   the Lagrangian equations of motion  determine them in terms of the physical fields as
their Hodge duals.
  Since the fields $y^m$ have to be included for a correct description of the
dynamics of the theory, it is convenient to adopt a complex notation also for the vector-tensor sector and work
with holomorphic coordinates $z^m \equiv z^m(P_M, y^m)$ together with their complex conjugates $\ol z^\mb$ (with
$m,\mb=1,\cdots , n_T$).  Using this notation, it is natural to introduce a collective holomorphic world-index
$i=(r,m)=1,\cdots , n_V+n_T$ on the  $2(n_V+n_T)$-dimensional embedding manifold $\cM_{(emb)}$, in parallel to
what has been done for gauge indices. This notation is quite natural from the point of view of dualization of
the standard supergravity theory, where $z^m$ are part of the K\"ahler coordinates $z^i$. According to it, we
will extend the set of spinors $\lambda^r$ to $\lambda^i$, including among them the spinors $\chi_M$, such that
$\chi_{M}^A= p_{Mi}\lambda^{iA}$, $\chi_{MA}=\bar p_{M\ib}\lambda^\ib_A$. Using the collective index formalism
the theory will look quite like the standard $N=2$ supergravity coupled to vector multiplets only, and this
explains, as we will see in the following, that most of the results coming from Bianchi Identities will look
formally like those of the standard $N=2$ supergravity, or a suitable extension of it. Since then the Free
Differential Algebra involves both the antisymmetric tensors $B_M$ and the degrees of freedom $y^m$, we expect
that the closure of the superspace Bianchi identities should imply the duality relation between them. In fact
this is what happens, see eq. \eq{dualization}, implying that the dualization relation is valid only on-shell.
As a consequence, the on-shell geometry will look rather different from its off-shell counterpart. In
particular, in the absence of a factorization of the two $\sigma$-models $\cM_T$ and $\cM_V$, the off-shell
K\"ahler--Hodge structure is completely destroyed since the metric is not even hermitian.

We finally note that, exactly like in the five dimensional case, the massive vector-tensor multiplets of the
$N=2$ four dimensional theory are short, BPS multiplets. This is in a contrast with what happens for the
scalar-tensor multiplets, where the tensor field is Hodge-dual to a scalar in the hypermultiplet sector
\cite{D'Auria:2004yi}. In that case, the multiplet becomes massive by introducing an appropriate coupling to a
vector multiplet. In our case, instead, the degrees of freedom corresponding to $A^M$ and $y^m$ do not have
spinor partners, but act as bosonic Lagrange multipliers in the theory.
 Being BPS multiplets, they are therefore charged and this in turn requires for CPT invariance that the vector-tensor multiplet
sector always includes an even number of tensor fields.

\section{The structure of the bosonic Free Differential Algebra}\label{bosfda}
Let us summarize here the main properties of the bosonic Free Differential Algebra underlying the supergravity
theory, found in \cite{mip}. \footnote{With an abuse of notation, we will call here with the same name $H_M$,
$F_M$ and $F^\Lambda$ the bosonic field-strengths associated to the corresponding forms on superspace.} It
reads:
\begin{equation} \left\{ \begin{array}{rcl}
F^\Lambda &=& dA^\Lambda + \frac 12 f_{\Sigma\Gamma}{}^\Lambda A^\Sigma\wedge A^\Gamma + m^{\Lambda M}B_M\\
F_M &=& d A_M + \hat T_{\Lambda M}{}^N A^\Lambda \wedge A_N\\
H_M &=& dB_M + T_{\Lambda M}{}^N A^\Lambda \wedge B_N +\left(d_{\Lambda\Sigma M} A^\Sigma +\hat T_{\Lambda M}{}^N A_N\right)\wedge F^\Lambda
\end{array} \right.
\label{fda}\end{equation}
 and it closes the Bianchi identities
\begin{equation} \left\{ \begin{array}{rcl}
 \nabla F^\Lambda &=&  m^{\Lambda M} H_M\\
 \nabla  F_M &=& 0\\
 \nabla H_M &=& \left(d_{\Lambda\Sigma M} F^\Sigma +\hat T_{\Lambda M}{}^N F_N\right)\wedge F^\Lambda
\end{array} \right.\end{equation}
where the  covariant derivatives are defined as follows:
\begin{eqnarray}
 \nabla F^\Lambda &\equiv & d F^\Lambda + \hat f_{\Lambda\Sigma}{}^\Gamma A^\Sigma \wedge F^\Gamma + m^{\Lambda M}\hat T_{\Sigma M}{}^N A_N \wedge F^\Sigma \label{hatfl}\\
 \nabla F_M &\equiv & d F_M + \hat T_{\Lambda M}{}^N \left(A^\Lambda \wedge F_N  - A_N \wedge F^\Lambda\right)\label{hatfm}\\
 \nabla H_M &\equiv & dH_M + \hat T_{\Lambda M}{}^N A^\Lambda \wedge H_N\label{hath}
\end{eqnarray}
provided  the following  constraints are satisfied:
\begin{eqnarray}
 f_{[\Sigma\Gamma}{}^\Delta f_{\Lambda]\Delta}{}^\Pi &=&0\label{bbi1}\\
 T_{[\Lambda M}{}^P  T_{\Sigma ]P}{}^N &=&\frac 12 f_{\Lambda\Sigma}{}^\Gamma T_{\Gamma M}{}^N\label{bbi2}\\
 T_{\Lambda M}{}^N &=& d_{\Lambda\Sigma M}m^{\Sigma N}\label{bbi3}\\
 m^{\Lambda (P}T_{\Lambda M}{}^{N)} &=&0\label{bbi4}\\
 f_{\Sigma\Gamma}{}^\Lambda m^{\Gamma M}&=& m^{\Lambda N}T_{\Sigma N}{}^M\label{bbi5}\\
 \hat T_{[\Lambda |M}{}^N d_{\Gamma|\Sigma]N}&-& f_{[\Lambda | \Gamma}{}^\Delta  d_{\Delta |\Sigma] M}
  -\frac 12 f_{\Lambda  \Sigma}{}^\Delta d_{\Gamma\Delta M}=0\label{bbi6}
 \end{eqnarray}
In the relations above we used the definition:
\begin{eqnarray}
 \hat T_{\Lambda M}{}^N&\equiv & T_{\Lambda M}{}^N + d_{\Sigma\Lambda M}m^{\Sigma N}= 2
 d_{(\Lambda\Sigma )M} m^{\Sigma N}\label{that}\\
 \hat f_{\Sigma\Gamma}{}^\Lambda &\equiv& f_{\Sigma\Gamma}{}^\Lambda + d_{\Gamma\Sigma M}
 m^{\Lambda M}\label{fhat}
 \end{eqnarray}
From \eq{bbi1}-\eq{bbi6} and \eq{that}, \eq{fhat} we derive the further, useful relations (see \cite{mip}):
\begin{eqnarray}
\hat f_{\Lambda\Gamma}{}^\Delta \hat f_{\Sigma \Delta}{}^\Pi - \hat f_{\Sigma\Gamma}{}^\Delta \hat f_{\Lambda \Delta}{}^\Pi&=&-\hat f_{\Lambda\Sigma}{}^\Delta \hat f_{ \Delta\Gamma}{}^\Pi\label{hatbbi1}\\
\hat T_{[\Lambda M}{}^P \hat T_{\Sigma ]P}{}^N &=&\frac 12 f_{\Lambda\Sigma}{}^\Gamma \hat T_{\Gamma M}{}^N\label{hatbbi2}\\
 m^{M (P}\hat T_{\Sigma M}{}^{N)} &=&0\label{hatbbi3}\\
\hat  f_{\Sigma\Gamma}{}^\Lambda m^{\Gamma M}&=& m^{\Lambda N}\hat T_{\Sigma N}{}^M\label{hatbbi4}\\
\hat  f_{\Sigma\Gamma}{}^\Lambda m^{\Sigma M}&=&0\label{hatbbi5}\\
 m^{\Lambda P}\hat T_{\Lambda M}{}^{N} &=&0\label{hatbbi6}
 \end{eqnarray}
Let us observe, following \cite{mip}, that the Free Differential Algebra written above contains, in the
definition of the field strengths, gauge couplings  different from the ones in the Bianchi identities
(non-``hatted'' versus ``hatted'' couplings), that is the couplings in the covariant derivatives and Bianchi
identities are deformed with respect to the defining ones. This is an unavoidable peculiarity of our request
\eq{bbi1} of closure of the gauge algebra Jacobi identities, a consistency condition that must be satisfied by
the gauge algebra $G$ if we require it to be an electric algebra. However, as shown in \cite{mip},  via the
field redefinition
$$B_M \to B_M + d_{\Lambda\Sigma M}A^\Lambda\wedge A^\Sigma$$
the Free Differential Algebra can be put in a form where only the hatted couplings appear. In this new,
equivalent setting, however, closure of the gauge algebra underlying the Free Differential Algebra is not
manifest, since the Jacobi identities are not satisfied. This is the setting generally used in \cite{de
Wit:2007mt}. Note that in \cite{de Wit:2007mt}, where the Free Differential Algebra was studied in the embedding
tensor framework, the closure of the algebra was however guaranteed by the fact that the embedding tensor should
satisfy a set of quadratic constraints, that are precisely the same as \eq{bbi1}-\eq{bbi6}.

%%%%%%%%%%%%%%%%%%%%%%%%%%%%%%%%%%%%%%%%%%%%%%% %%%%%%%%%%%%%%%%%%%%%%%%%%%%%%%%%%%%%%%%%%%%%%

\subsection{Gauge invariance properties and symplectic embedding}\label{emb}
Let us study in more detail the gauge structure of the Free Differential Algebra \eq{fda}.
\subsubsection{2-form gauge transformation}
Eq. \eq{fda} is invariant under the 2-form gauge transformation with 1-form parameter $\Lambda_M$:
 \begin{equation} \left\{ \begin{array}{rcl}
\delta B_M &=& d\Lambda_M + T_{\Lambda M}{}^N A^\Lambda \wedge \Lambda_N \equiv D\Lambda_M
\\
\delta A^\Lambda &=& - m^{\Lambda M}\Lambda_M\\
\delta A_M &=& 0
\end{array} \right.\,,
\label{lambdagauge}\end{equation}
under which
\begin{equation} \left\{ \begin{array}{rcl}
\delta H_M &=& 0
\\
\delta F^\Lambda &=& 0\\
\delta F_M &=& 0
\end{array} \right.\,.
\label{lambdagaugecurv}\end{equation}
However, we still have the freedom to redefine $B_M \to B_M + k_{\Lambda\Sigma M}A^\Lambda\wedge A^\Sigma$ for a generic constant tensor $k_{\Lambda\Sigma M}$. We
can exploit this freedom to fix the 1-form gauge $\bar \Lambda_M$ such that:
\begin{equation} \left\{ \begin{array}{rcl}
A^\Lambda &\to & A'^\Lambda =\delta^\Lambda_X A^X- m^{\Lambda M}\bar\Lambda_M\\
 A_M &\to& A'_M= A_M\\
 B_M &\to & B_M'= \bar B_M +d\bar \Lambda_M +T_{XM}{}^NA^X \bar \Lambda_N+\nn\\
 &&-\frac 12 d_{XY M} A^X
\wedge A^Y -\frac 12 d_{\Lambda\Sigma M} m^{\Lambda N} m^{\Sigma P}\bar \Lambda_N
\wedge \bar\Lambda_P
\end{array} \right.\,,
\label{lambdagaugegf}\end{equation}
the free differential agebra turns out to be written only in  terms of physical massive fields:
\begin{equation} \left\{ \begin{array}{rcl}
 F'^\Lambda &= & \delta^\Lambda_X F^X+ m^{\Lambda M}\bar B_M\\
 F'_M &=& F_M\\
H'_M &=& d\bar B_M +\hat T_{XM}{}^N\left(A^X \bar B_N +A_N F^X\right) +d_{(XY)M}A^X F^Y+\nonumber\\
&&+ \left(f_{XY}{}^W d_{[ZW] M} +T_{XM}{}^N d_{YZ N}\right) A^X \wedge A^Y \wedge A^Z
\end{array} \right.\,,
\label{lambdagaugecurvgf}\end{equation}

\subsubsection{1-form gauge transformations and symplectic embedding}
The Free Differential Algebra is also covariant under the 1-form gauge transformation with parameters
$\epsilon^\Lambda$, $\epsilon_M$:
\begin{equation}
\left\{ \begin{array}{rcl}
\delta A^\Lambda &=& d\epsilon^\Lambda +   f_{\Sigma\Gamma}{}^\Lambda A^\Sigma \epsilon^\Gamma \\
\delta A_M &=& d \epsilon_M + \hat T_{\Lambda M}{}^N \left(A^\Lambda  \epsilon_N - A_N\epsilon^\Lambda\right)\\
\delta B_M &=& - T_{\L M}{}^N \e^\L B_N -
\left( d_{\L\S M} \e^\S +\hat T_{\L M}{}^N \e_N \right) F^\L
\end{array} \right.\,,
\label{gaugetransf}\end{equation}
under which:
\begin{equation}
\left\{ \begin{array}{rcl}
\delta F^\Lambda &=&  -\hat f_{\Gamma\Sigma}{}^\Lambda F^\Sigma \epsilon^\Gamma - m^{\Lambda N} \hat T_{\Sigma N}{}^M F^\Sigma
\epsilon_M  \\
\delta F_M &=& \hat T_{\Lambda M}{}^N \left( F^\Lambda \epsilon_N - F_N \epsilon^\Lambda \right) \\
\delta H_M &=& - \hat T_{\Lambda M}{}^N \epsilon^\Lambda H_N
\end{array} \right.\,.
\label{gaugetransfs}\end{equation} Let us emphasize that, as explained in Section 2, the theory contains
electric  gauge vectors $A^\Lambda=(A^X,A^M)$ together with magnetic ones $A_M$, since all of them are needed to
implement the Higgs mechanism giving mass to the vector-tensor multiplet. In particular, as we will see in the
following, the equations of motion of the $B$-fields identify, on-shell, the field-strengths $F_M=
\cN_{M\Lambda}F^{+\Lambda}+ \ol\cN_{M\Lambda} F^{-\Lambda}$ (see eq. \eq{fmlambda0}) with the magnetic
field-strengths $\cG_M \equiv -\frac 12 {}^*\frac{\d}{\d F^M}(\cL_k+\cL_{Pauli})$, where $\cL_k$ and
$\cL_{Pauli}$ are the kinetic and Pauli Lagrangians respectively. Then, we can write a symplectic vector of
electric and magnetic field strengths as:
$$\cF^\alpha =\left(F^\L, \cG_X,F_M\right)\,,$$ with $\alpha=1,\cdots ,2(1+n_V+n_T)$ being a
 symplectic index running over all the electric and magnetic fields.
 $\cG_X \equiv -\frac 12 {}^*\frac{\d}{\d F^X}(\cL_k+\cL_{Pauli})$ is the magnetic field-strength dual to $F^X$.
Then, the gauge variations of the field-strenghts, eq. \eq{gaugetransfs}, can be written as:
\begin{equation}
\delta_\epsilon \cF^\alpha =-\cF^\gamma \left(\cT_\beta\right)_\gamma{}^\alpha \epsilon^\beta
\end{equation}
where $\cT_\alpha$ are the gauge algebra generators embedded in the symplectic group. They can be written in
block form as
\begin{equation}
(\cT_\alpha)_\beta{}^\gamma =
\begin{pmatrix}
(\cT_\alpha)_\S{}^\G & (\cT_\alpha)_{\S\G} \\
(\cT_\alpha)^{\S\G} & (\cT_\alpha)^\S{}_\G
\end{pmatrix}
\end{equation}
where, for $\cT$ to be symplectic,
\begin{equation}(\cT_\alpha)^\S{}_\G = - (\cT_\alpha)_\S{}^\G, \quad (\cT_\alpha)_{\S\G} =
(\cT_\alpha)_{\G\S}, \quad (\cT_\alpha)^{\S\G} =
(\cT_\alpha)^{\G\S}\label{symplectic}
\end{equation}
Condition \eq{symplectic}, together with \eq{gaugetransfs}, allows to completely determine the $\cT_\alpha$:
\begin{eqnarray}
(\cT_\Lambda)_\beta{}^\gamma &=&
\begin{pmatrix}
\hat f_{\Lambda\Sigma}{}^\G & (\cT_\Lambda)_{XY}\delta_{(\S}^X\delta_{\G )}^Y \\
0 & -\hat f_{\Lambda\Gamma}{}^\Sigma
\end{pmatrix}
\label{tlambda}
\\
(\cT^P)_\beta{}^\gamma &=&
\begin{pmatrix}
m^{\Gamma N}\hat T_{\Sigma N}{}^P & -2 \delta_{(\S }^N\hat T_{\G ) N}{}^P+ (\mathbf{T}^P)_{XY}\delta_{(\Lambda}^X\delta_{\Sigma )}^Y \\
0 & -m^{\Sigma N}\hat T_{\Gamma N}{}^P
\end{pmatrix}
\label{tp}\\
 (\cT^X)_\beta{}^\gamma &=&0\label{tx}
 \end{eqnarray}
 Note that the tensor $(\cT_\Lambda)_{XY}$  is not relevant in the vector-tensor sector of the theory,
 as it is not coupled to fields in the tensor multiplets, however it corresponds to a possible deformation
 of the gauging
 in the vector multiplet directions, for gaugings having a non-homogeneous action on the vector-kinetic matrix
  \cite{de Wit:1984px}.
 The symplectic embedding above allows to predict the form of the gauge transformation of the magnetic
  field-strengths $\cG_X$:
\begin{eqnarray}
\delta \cG_X&=& -\left(\cT_{\Lambda XY}\epsilon^\Lambda +(\cT^P)_{ XY}\epsilon_P\right)\,
F^Y + \hat T_{XN}{}^P\,\epsilon_P\,F^N+\nonumber\\
&&+ f_{\Lambda X}{}^Y\,\epsilon_\Lambda\,\cG_Y +\left(\hat  f_{\Lambda X}{}^N\,\epsilon_\Lambda +
 \hat T_{XM}{}^P\,m^{NM}\,\epsilon_P\right)\,F_N\,.
\end{eqnarray}

 The  generators $\cT_\alpha$ have to satisfy a set of relations corresponding to a
 no-anomaly condition \cite{DeRydt:2008hw}, which is easily expressed in the embedding tensor formalism and reads:
\begin{equation}
(\cT_{(\alpha})_\beta{}^\delta \,\IC_{\gamma
)\delta}=0\,.\label{noanomaly}
\end{equation}
It corresponds to the following relations:
\begin{eqnarray}
(\cT_{(\Lambda})_{\Sigma\Gamma)}&=&0\label{anomaly1}\\
(\cT^{\Gamma})_{\Lambda\Sigma}&=&2 (\cT_{(\Lambda})_{\Sigma)}{}^\Gamma\label{anomaly2}\\
(\cT_{\Gamma})^{\Lambda\Sigma}&=&-2 (\cT^{(\Lambda})_{\Gamma}{}^{\Sigma)}\label{anomaly3}
\\
(\cT^{(\Lambda})^{\Sigma\Gamma)}&=&0\label{anomaly4}
\end{eqnarray}
In particular, eq. \eq{anomaly1} implies
\begin{equation}
(\cT_M)_{XY}=0\,,
\end{equation}
eq. \eq{anomaly2}, using  \eq{that}, implies
  \begin{eqnarray}
m^{\Gamma M}d_{(\Lambda\Sigma\Gamma )}&=&\frac 16 (\mathbf{T}^M)_{XY}\delta_{(\Lambda}^X\delta_{\Sigma)}^{Y}
\,,\label{dsymm}\\
m^{X M}\,d_{(\Lambda N)M}&=&0\label{mxp}\,.
  \end{eqnarray}
  Eq. \eq{dsymm}\ in turns implies, by multiplication with $m^{\Sigma N}$
  \begin{eqnarray}
m^{M (P}\hat T_{\Sigma M}{}^{N)} &=&0\,,\label{hatbbi3'}
  \end{eqnarray}
  while eq. \eq{mxp} is trivially satisfied if we choose, as we will do throughout the paper, a basis where \begin{equation}
m^{XP}=0\quad\,, \quad \mbox{det}[m^{MP}]\neq 0\,.\label{mbasis}
\end{equation}

  Eq. \eq{anomaly3}, using \eq{bbi4} and  \eq{hatbbi3'}, implies
  \begin{eqnarray}
(\cT_{\Gamma})^{\Lambda\Sigma}&=&0\,,
  \end{eqnarray}
  while eq. \eq{anomaly4} is trivially satisfied.
  Let us finally observe that the following relation holds:
      $(\cT_\Lambda)_\alpha{}^\beta m^{\Lambda M}=0$ as is easily checked using \eq{bbi1} - \eq{bbi6}.
      In the basis \eq{mbasis}, it corresponds to the statement $\cT_M=0$.

The $\cT_\alpha$, subject to \eq{bbi1} - \eq{bbi6} together with \eq{anomaly1} - \eq{anomaly4},
   close the algebra in the symplectic representation:
 \begin{equation}
\left[\cT_{\alpha},\cT_\beta \right] = -\cT_{\alpha\beta}{}^\gamma \,\cT_\gamma\,.\label{embalg}
\end{equation}
This will be shown in Appendix \ref{algebraemb}.

Note that the eqs. \eq{anomaly1} - \eq{anomaly4} are necessary
conditions to have gauge and supersymmetry invariance of the $N=2$
Lagrangian including topological terms of generalized Chern--Simons
type. The symplectic embedding built up above is crucial  to show
this property.  Indeed, let us recall that in the gauged theory,
according to \cite{Gaillard:1981rj},  the gauge group has to be
embedded in the symplectic group, as we found above (see
\eq{tlambda}, \eq{tp}). We generally have, for an infinitesimal
electric symplectic rotation $S=\iden - s$ with
$s=\begin{pmatrix}a&c\cr 0& -a^t\end{pmatrix}$  of the (self-dual
part of the) field strength $(\cF^-)^\alpha$:
\begin{equation}
(\cF'^-)^\beta = (\cF^-)^\alpha S_\alpha{}^\beta
\end{equation}
implying, recalling that $\cG^-_\Lambda = \ol \cN_{\Lambda\Sigma}(F^-)^\Sigma$
\begin{equation}
\delta \cN = -c + a^t \cN + \cN a\,.
\end{equation}
In extended supergravity the gauge algebra has to be embedded in the symplectic algebra. In particular, for an
electric theory the infinitesimal gauge transformations are given by a matrix $s$ where
\cite{Andrianopoli:2002aq}:
\begin{equation}
s_\alpha{}^\beta = \epsilon^\Lambda \left(\cT_\Lambda\right)_\alpha{}^\beta
\end{equation}
For theories including magnetic vectors, the above relation generalizes to \cite{deWit:2005ub,de Wit:2007mt}
\begin{equation}
s_\alpha{}^\beta = \epsilon^\Lambda \left(\cT_\Lambda\right)_\alpha{}^\beta + \epsilon_\Lambda
\left(\cT^\Lambda\right)_\alpha{}^\beta
\end{equation}
This implies that, when the constant matrix $c$ is different from zero, the supersymmetric and gauge invariant
Lagrangian must include a generalized Chern--Simons topological term \cite{de Wit:1984px}. In particular, in our
case we have
\begin{equation}
c_{\Lambda\Sigma}=\epsilon^\Gamma( \cT_\Gamma)_{\Lambda\Sigma}+\epsilon_M( \cT^M)_{\Lambda\Sigma}
\end{equation}
 It is worth noticing that the  gauge invariance of the Lagrangian requires
 \eq{anomaly1} - \eq{anomaly4}, together with the further conditions  (antisymmetrization in $\Lambda_1\Lambda_2\Lambda_3\Lambda_4$
  is understood in \eq{che}):
  \begin{eqnarray}
&&  f_{[\Lambda\Sigma}{}^\Omega (\cT_{\Gamma]})_{\Delta\Omega}=f_{[\Lambda\Sigma}{}^\Omega (\cT_{\Gamma})_{\Delta]
\Omega}\label{dlv}\\
&&f_{\Lambda_1 \Lambda_2}{}^\Theta \Bigl[f_{\Theta\Delta}{}^\Omega (\cT_{\Lambda_4})_{\Lambda_3\Omega}-
f_{\Lambda_3\Lambda_4}{}^\Omega (\cT_{\Delta})_{\Theta\Omega} +
f_{\Delta\Lambda_4}{}^\Omega (\cT_{\Lambda_3})_{\Theta\Omega}-\nonumber\\
&&\hskip 1.3cm +f_{\Delta\Lambda_4}{}^\Omega (\cT_{\Omega})_{\Lambda_3\Theta}\Bigr]=0 \label{che}
  \end{eqnarray}
These relations do  not involve the tensor sector, since they only include electric couplings. It is anyway
interesting to analyze them in more detail, in the context of the gaugings introduced in \cite{de Wit:1984px}.
In fact, eq. \eq{dlv} coincides with the condition (3.18) in  \cite{de Wit:1984px}. As far as eq. \eq{che} is
concerned, actually it has a geometric meaning since it corresponds to a cohomological statement. Indeed, let us
consider the tensor $t_{\Lambda\Sigma\Gamma\Delta}\equiv f_{[\Lambda\Sigma}{}^\Omega
(\cT_{\Gamma]})_{\Delta\Omega}
 =-\frac 12 t_{[\Lambda\Sigma\Gamma\Delta]}$.
 Eq. \eq{che} can be rewritten, using \eq{dlv}, as:
\begin{equation}
  3t_{\Sigma\Gamma[\Lambda_1\Lambda_2}f_{\Lambda_3\Lambda_4]}{}^\Gamma +
  2f_{\Sigma[\Lambda_1}{}^\Gamma t_{\Lambda_2\Lambda_3\Lambda_4]\Gamma}=0\label{cheval}
 \end{equation}
  This equation has a simple interpretation in terms of the Chevalley--Eilenberg Lie algebra cohomology of the
 gauge group $G$ \cite{chevalley}.
Indeed, the free differential
 algebra \eq{fda} is constructed starting from the algebra of the gauge group, which in dual form is expressed
 by the Cartan--Maurer equation \cite{D'Auria:1980ra}:
 \begin{equation}
 dA^\Lambda + \frac 12 f_{\Sigma\Gamma}{}^\Lambda A^\Sigma\wedge A^\Gamma =0\,.
 \end{equation}
We recall that for a generic $p$-form in a given representation $D(\mathbf{T}_\Lambda)$ of $G$ labeled by the
index $K$
$$\omega^{(p)}_K=\omega^{(p)}_{K|\Lambda_1 \cdots\Lambda_p} A^{\Lambda_1}\wedge \cdots \wedge A^{\Lambda_p}$$
the condition for $\omega^{(p)}_K$ to be a representative of a cohomology class $H^{(p)}$ of the Lie algebra is
$$\partial \omega^{(p)}_K=0\,$$ where
\begin{equation}
\partial \omega^{(p)}_K \equiv \nabla \omega^{(p)}_K = -\frac 12 f_{\Sigma\Gamma}{}^\Lambda
A^\Sigma \wedge A^\Gamma \wedge  (i_\Lambda \omega^{(p)}_K) + D(\mathbf{T}_\Lambda)_K{}^L \wedge \omega^{(p)}_L
\end{equation}
and $i_\Lambda$ denotes contraction along the generator $\mathbf{T}_\Lambda$.
 Then if we consider $t_{\Lambda\Sigma\Gamma\Delta}$ as the component of a 3-form in the adjoint representation of $G$:
 \begin{equation}
 t_\Lambda \equiv t_{\Lambda\Sigma\Gamma\Delta} A^\Sigma\wedge A^\Gamma\wedge A^\Delta
 \end{equation}
  then eq. \eq{cheval} is just the condition that $t_\Lambda$ lies in the cohomology class
  $H^{(3)}$ of the Chevalley--Eilenberg cohomology of the
  $\{A^\Lambda\}$ \footnote{
 This is a general statement. Note that it  extends  eq. (3.17) of \cite{de Wit:1984px} also to theories
 where a prepotential $F(X)$ does not exist.}.
Notice moreover that also eq. \eq{bbi6} has a Lie algebra cohomology interpretation: indeed, introducing the
1-form $\Phi_{\L M} \equiv d_{\L\S M} A^\S$,  it can be easily verified that \eq{bbi6} corresponds the condition
for $\Phi_{\L M}$ to lie in the Chevalley--Eilenberg cohomology class $H^{(1)}$.

\section{The $N=2$ theory of supergravity coupled to vector and vector-tensor multiplets}\label{theory}
%%%%%%%%%%%%%%%%%%%%%%%%%%%%%%%%%%%%%%%%%%%%%%% %%%%%%%%%%%%%%%%%%%%%%%%%%%%%%%%%%%%%%%%%%%%%%
\subsection{Definition of Superspace Curvatures}
According to the geometric approach, we define the Free Differential Algebra of our theory as follows:
    \begin{itemize}
    \item
    We start with  the Free Differential Algebra of pure supergravity:
\begin{align} \cR^a{}_b &= \de \o^a{}_b - \o^a{}_c \wedge \o^c{}_b \label{curvature} \\ \mathcal{T}^a &= \de V^a -
\o^a{}_b V^b - \ii \ol \psi_A \g^a \psi^A \label{tors} \\ \rho_A &= \de \psi_A - \qu \o_{ab} \g^{ab} \psi_A +
\imez \mathcal{Q} \psi_A  \label{rhol} \\ \rho^A &= \de \psi^A - \qu \o_{ab} \g^{ab} \psi^A - \imez \mathcal{Q}
\psi^A \label{rhor} \,. \end{align} where with $\mathcal{Q}$ we denote a gauged $U(1)$ connection, which is
the remnant of the gauged $U(1)$-K\"ahler composite connection of special geometry. We recall that in the
 these definitions the spin connection $\omega^a{}_b$ and the bosonic and fermionic component of the
 supervielbein $V^a, \psi_A,\psi^A$, as well as $\cQ$, are superspace 1-forms , the left-hand sides
 definining the corresponding superspace curvatures.\footnote{For the definition of the gauged
 $U(1)$-K\"ahler composite connection of special geometry in terms of the ungauged one and for
  all the notation concerning special geometry, we refer the reader to the standard $N=2$, $D=4$
supergravity of ref. \cite{Andrianopoli:1996cm}.}
\item
To complete  the superspace Free Differential Algebra, the  bosonic space-time fields and curvatures
introduced  in Section \ref{bosfda} for the vector and tensor field-strengths are suitably generalized to
their superspace extension as follows:
\begin{align}
 F^\Lambda &= \de A^\Lambda + \frac 12 f_{\Sigma\Gamma}{}^\Lambda A^\Sigma A^\Gamma + m^{\Lambda M} B_M + L^\Lambda \ol \psi^A \psi^B \e_{AB} + \ol
L^\Lambda \ol \psi_A \psi_B \e^{AB} \label{Flambda} \\
F_M &= \de A_M + \hat T_{\Lambda M}{}^N A^\Lambda A_N + L_M \ol \psi^A
\psi^B \e_{AB} + \ol L_M \ol \psi_A \psi_B \e^{AB} \label{FM} \\
H_M &= \de B_M + T_{\Lambda M}{}^N
A^{\Lambda} B_N  + 2\ii P_M \ol \psi_A \g_a \psi^A V^a + \nn \\ &+ \left( d_{\Lambda\Sigma M} A^\Sigma + \hat T_{\Lambda M}{}^N A_N
\right) \left( F^\Lambda - L^\Lambda \ol \psi^A \psi^B \e_{AB} - \ol  L^\Lambda \ol \psi_A \psi_B \e^{AB} \right) \label{HM}
\end{align}
Here $P_M$ is a real section on the $\sigma$-model, while $L^\L$, $\ol L^\L$, $L_M$ and $\ol L_M$ are
complex sections on the $\sigma$-model, analogous to the covariantly holomorphic sections of special
geometry.
\item
 Finally, the Free Differential Algebra is enlarged to include the 1-form gauged field-strengths for the
0-form complex scalars $z^i$ and 0-forms spin $1/2$ spinors $\lambda^{iA}, \lambda^{\bar i}_{A}$ belonging
to the $N=2$ vector multiplet and tensor multiplet representations of supersymmetry:
\begin{align} \nabla z^i &= \de z^i + k^i_\L A^\L  -  k^{iM}A_M \label{zi} \\
\na \l^{iA} &= \de \l^{iA} - \qu \o_{ab} \g^{ab}
\l^{iA} - \imez \cQ \l^{iA} +  \G^i{}_j \l^{jA}
%+ \G^i{}_{\jb} \e^{AB} \l^{\jb}_B
\label{lambdai}
\end{align} where $\G^i{}_j= \G^i{}_{jk} \nabla z^k$ is the gauged K\"ahlerian Levi--Civita connection
$(1,0)$-form on the embedding $\sigma$-model $\cM_{(emb)}$. $k^i_\L$ are the complex Killing vectors in the adjoint representation
of the algebra $G$ while $k^{iM}$ are Killing vectors in the appropriate representation of $G_0$ (the
invariant subgroup of $G$ or, more generally, its contraction). This choice complies, in our redundant
formulation, to the requirement that the vectors $A_M$ undergo the Higgs mechanism by eating the real
degrees of freedom $Y^M$ dual to $B_M$. This will prove to be consistent with the solution of the superspace
Bianchi identities.

\end{itemize}

The construction of the theory, namely the supersymmetric
Lagrangian, its transformation laws and the constraints on the
$\sigma$-model,
 is obtained by working out the constraints obtained from the superspace Bianchi identities and/or
 the superspace equations of motion of the rheonomic Lagrangian thought of as a 4-form embedded in superspace.
 A short derivation and a summary of the results are given in   Appendices
  \ref{susyparam}, \ref{Lagrangian} and \ref{relations}. Restricting
 the rheonomic Lagrangian to the physical space-time
 we arrive at the following \emph{space-time Lagrangian}:

%%%%%%%%%%%%%%%

\begin{equation}
\cS = \int\sqrt{-g}\,d^4 x\,\left[\cL_k +\cL_{Pauli}+\cL_{gauge}+\cL_{4f}\right]
\end{equation}
where
\begin{align}
\cL_k =& \,-\frac 12 \cR + \left(\ol\psi^A_\mu \gamma_\sigma\rho_{A|\nu\rho}-\ol\psi_{A\mu}
\gamma_\sigma\rho^A_{\nu\rho}\right)\frac{\epsilon^{\mu\nu\rho\sigma}}{\sqrt{-g}}
%+\nonumber\\&\,
+\ii \left(\bar\cN_{\Lambda\Sigma} \tilde\cF^{-\Lambda}_{\mu\nu}\tilde\cF^{-\Sigma |\mu\nu}-\cN_{\Lambda\Sigma} \tilde\cF^{+\Lambda}_{\mu\nu}\tilde\cF^{+\Sigma |\mu\nu}\right)+\nonumber\\
&\,+\frac 1{16}  \cM^{MN}  Y_{M|\mu}Y_{N}^{\,\mu}
-\frac 18  \cM^{MN}Y_{M|\mu }\tilde H_{N |\nu\rho\sigma}\frac{\epsilon^{\mu\nu\rho\sigma}}{\sqrt{-g}}+\nn\\
&+\frac\ii 4\cM^{MN}\tilde H_{M|\mu\nu\rho}\left(p_{Ni}\tilde Z^i_\sigma -\bar p_{N\ib}
\tilde{\bar Z}^\ib_\sigma\right)\frac{\epsilon^{\mu\nu\rho\sigma}}{\sqrt{-g}}+\nonumber\\
&\,+\frac 12 G_{ij}\tilde Z^i_\mu\tilde Z^{j|\mu}+G_{i\jb}\tilde Z^i_\mu\tilde{ \bar Z}^{\jb|\mu}+\frac 12 G_{\ib\jb}\tilde{ \bar Z}^\ib_\mu\tilde{ \bar Z}^{\jb |\mu}
%+\nonumber\\&
-\frac \ii 2 g_{i\jb}\left(\ol\lambda^{iA}\gamma^\mu\nabla_\mu \lambda^\jb_A +\ol\lambda^\jb_A\gamma^\mu\nabla_\mu
\lambda^{iA}\right)\label{lkin}
\end{align}
\begin{align}
\cL_{Pauli} =& \,-\frac{\epsilon^{\mu\nu\rho\sigma}}{\sqrt{-g}}\left(\cF^\Lambda_{\mu\nu}+L^\Lambda \ol\psi^A_{[\mu}\psi^B_{\nu]}\epsilon_{AB}+\ol L^\Lambda \ol\psi_{A[\mu}\psi_{B|\nu]}\epsilon^{AB}\right)\times\nonumber\\
&\Bigl[\cN_{\Lambda\Sigma}L^\Sigma \ol\psi^A_{[\rho}\psi^B_{\sigma]}\epsilon_{AB}+\ol  \cN_{\Lambda\Sigma}\ol L^\Sigma \ol\psi_{A[\rho}\psi_{B|\sigma]}\epsilon^{AB}+\nonumber\\
&-\ii\ol  \cN_{\Lambda\Sigma} f^\Sigma_i
\ol \psi^A_{[\rho} \g_{\sigma]} \l^{i B} \e_{AB}   - \ii \cN_{\Lambda\Sigma}\ol  f^\Sigma_{\ib}
\ol \psi_{A[\rho} \g_{\sigma]} \l^{\ib}{}_B \e^{AB}+\nonumber\\
&-\frac \ii 4\left(X_{\Lambda ij}\ol\lambda^{iA}\gamma_{\rho\sigma}\lambda^{jB}\epsilon_{AB}+\ol X_{\Lambda \ib\jb}\ol\lambda^{\ib}_A\gamma_{\rho\sigma}\lambda^{\jb}_B\epsilon^{AB}\right)\Bigr]+\nonumber\\
&+g_{i\jb}\left(\nabla_\mu z^i \ol\lambda^{\jb}_A\gamma^{\mu\nu}\psi^A_\nu +\nabla_\mu \ol z^\jb \ol\lambda^{iA}\gamma^{\mu\nu}\psi_{A\nu}\right)+\nonumber\\
&-Q^M \left(p_{Mi}\nabla_\mu z^i -\bar p_{M\ib}\nabla_\mu \bar z^\ib\right)\ol\psi^A_\nu\gamma_\sigma\psi_{A|\rho}\frac{\epsilon^{\mu\nu\rho\sigma}}{\sqrt{-g}}+\nonumber\\
&+\frac \ii 2 \left[(\nabla_k g_{i\jb}-Q^M p_{Mk}g_{i\jb}) \nabla_\mu z^k -
(\nabla_\kb g_{i\jb}-Q^M \bar p_{M\kb}g_{i\jb}) \nabla_\mu \ol z^\kb\right]\ol\lambda^{iA}\gamma^\mu\lambda^\jb_A
\end{align}
\begin{eqnarray}
\cL_{gauge}= \cL_{mass}+\cL_{CS}-V(z,\ol z)
\end{eqnarray}
\begin{align}
\cL_{mass}=&\left[2S_{AB}\ol\psi^A_\mu\gamma^{\mu\nu}\psi^B_\nu +\ii g_{i\jb}W^{iAB}\ol\lambda^\jb_A\gamma_\mu\psi_B^\mu+ \cM_{iAjB}\ol\lambda^{iA}\lambda^{jB}\right]+\mbox{ h.c.}
\end{align}
 \begin{eqnarray}
\mathcal{L}_{CS} &=&  - m^{MN} \cF_M \wedge\left(B_N + \frac 12 d_{\Lambda\Sigma N} A^\Lambda A^\Sigma\right) +\nn\\
&&- m^{MN}d_{(\Lambda\Sigma ) N} \left(\cF^\Lambda -m^{\Lambda P}B_P\right)\wedge A_M \wedge\left(A^\Sigma +
 m^{\Sigma Q}A_Q\right) +\nonumber\\
 &&+\frac 13 ({\cT}_{\Lambda})_{\Sigma\Gamma} A^\Lambda \wedge A^\Sigma\wedge
 \left(\cF^\Gamma -m^{\Lambda M} B_M-\frac 18 f_{\Delta\Omega}{}^\Gamma A^\Delta \wedge A^\Omega\right) +\nn\\
&&-\frac 12 m^{NP}\left(d_{\Lambda\Sigma N}\hat T_{\Gamma M}{}^N -d_{[\Delta\Sigma ] N}f_{\Gamma\Lambda}{}^\Delta\right)
A^\Lambda \wedge A^\Sigma\wedge A^\Gamma \wedge A_P
  \end{eqnarray}
  \begin{align}
  V(z,\bar z) =& \, g_{i\jb}\left(k^i_\Lambda \ol L^\Lambda - k^{iM} \ol L_M\right)\left(\ol k^\jb_\Sigma  L^\Sigma
   - \ol k^{\jb N}  L_N\right)
  \end{align}
 \begin{align}
&\cL_{4f} = \frac{1}{2} \left( L^\L \ol \psi^A_\mu \psi^B_\nu \e_{AB} + \ol L^\L \ol \psi_{A \mu} \psi_{B\nu} \e^{AB} \right) \left( \cN_{\L\S} L^\S \ol \psi^C_\rho \psi^D_\s \e_{CD} + \ol \cN_{\L\S} \ol L^\S \ol\psi_{C\rho} \psi_{D\s}\e^{CD} \right) \frac{\e^{\mu\nu\rho\s}}{\sqrt{-g}} + \nn \\
& \hskip -3mm - \frac{1}{2} \left(f^\L_i \ol \psi^A_\mu \g_\rho \l^{iB} \e_{AB} + \bar f^\L_{\ib} \ol \psi_{A\mu} \g_\rho \l^\ib_B \e^{AB} \right) \left( \ol \cN_{\L\S} f^\S_j \ol \psi^C_\nu \g_\s \l^{jD} \e_{CD} + \cN_{\L\S} \bar f^\S_{\jb} \ol \psi_{C\nu} \g_\s \l^\jb_D \e^{CD} \right) \frac{\e^{\mu\nu\rho\s}}{\sqrt{-g}} + \nn \\
& - \frac{\ii}{2} g_{i\jb} \ol \l^{iA} \g_\rho \l^\jb_B \ol \psi_{A\mu} \g_\s \psi^B_\nu \frac{\e^{\mu\nu\rho\s}}{\sqrt{-g}} + \nn \\
& - \frac{1}{4} \left( f^\L_i \ol X_{\L\jb\kb} \ol \psi^A_\mu \g_\nu \l^{iB} \e_{AB} \ol \l^\jb_C \g^{\mu\nu} \l^\kb_D \e^{CD} - \bar f^\L_{\ib} X_{\L jk} \ol \psi_{A\mu} \g_\nu \l^\ib_B \e^{AB} \ol \l^{jC} \g^{\mu\nu} \l^{kD} \e_{CD} \right) + \nn \\
& - \frac{1}{6} \left( C_{ijk} \ol \l^{iA} \g^\mu \psi^B_\mu \ol \l^{jC} \l^{kD} \e_{AC} \e_{BD} - \ol C_{\ib\jb\kb} \ol \l^\ib_A \g_\mu \psi_B^\mu \ol \l^\jb_C \l^\kb_D \e^{AC} \e^{BD} \right) + \nn \\
& \hskip -15mm + \frac{1}{12}
\Bigl\{ \frac{3\ii}{16}(\cN -\ol\cN)_{\L\S} \left( C_{jkn}C_{i\ell m}g^ {m\mb}g^ {n\nb} \bar f^\L_\mb \bar f^\S_\nb\ol \l^{jA} \g_{ab} \l^{kB} \, \ol \l^{iC} \g^{ab} \l^{\ell D} \, \e_{AB} \e_{CD} + h.c.\right) + \nn \\
&  \hskip -5mm  -\ii  \Bigl[ \left( \nabla_i  C_{jkl} + 2 Q^M p_{M i} C_{jkl} + 3 C^m{}_{ij} C_{k\ell m}  \right) \ol \l^{iA} \l^{jC} \ol \l^{kB} \l^{\ell D} \e_{AB} \e_{CD} + h.c.\Bigr] + \nn \\
&  \hskip -5mm + 3\bigl[R_{i\jb k\lb}  - \frac{3}{2} g_{i\jb} g_{k\lb} +
 g_{i\lb} g_{k\jb}-\frac 12 g_{\ell\jb}\nabla_\lb C^\ell{}_{ik} -\frac 12 g_{k\kb}\nabla_i C^\kb{}_{\jb\lb}\bigr]\ol \l^{iA} \l^{kB} \ol \l^\jb_A \l^\lb_B  \Bigr\}
\end{align}
In writing the kinetic terms of the Lagrangian we have denoted with
a tilde the supercovariant field strengths defined as:
\begin{align}
\tilde\cF^{ \Lambda}_{\mu\nu}\equiv &\cF^{ \Lambda}_{\mu\nu}+L^\Lambda \ol\psi^A_{[\mu}\psi^B_{\nu]}\epsilon_{AB}+\ol L^\Lambda \ol\psi_{A[\mu}\psi_{B|\nu]}\epsilon^{AB}+\nonumber\\
&-\ii f^\L_i
\ol \psi^A_{[\mu} \g_{\nu]} \l^{i B} \e_{AB}   - \ii \ol  f^\L_{\ib}
\ol \psi_{A[\mu} \g_{\nu]} \l^{\ib}{}_B \e^{AB}\label{supcovflambda} \\
\tilde\cF_{ M|\mu\nu}\equiv &\cF_{M|\mu\nu}+L_M \ol\psi^A_{[\mu}\psi^B_{\nu]}\epsilon_{AB}+\ol L_M \ol\psi_{A[\mu}\psi_{B|\nu]}\epsilon^{AB}+\nonumber\\
&-\ii f_{Mi}
\ol \psi^A_{[\mu} \g_{\nu]} \l^{i B} \e_{AB}   - \ii \ol  f_{M\ib}
\ol \psi_{A[\mu} \g_{\nu]} \l^{\ib}{}_B \e^{AB}\label{supcovfm} \\
\tilde H_{M|\mu\nu\rho} \equiv& H_{M|\mu\nu\rho}+2\ii P_M \ol\psi^A_{[\mu}\gamma_\rho\psi^A_{\nu]}- p_{Mi} \ol
\psi_{A[\mu} \g_{\nu\rho]} \l^{iA}- \bar p_{M\ib} \ol \psi^A_{[\mu}
\g_{\nu\rho]} \l^{\ib}{}_A    \label{supcovHM}\\
\tilde Z^i_\mu \equiv & \nabla_\mu z^i -\ol\lambda^{iA}\psi_{A\mu}\label{supcovz}\\
\tilde{\bar Z}^\ib_\mu \equiv & \nabla_\mu \bar z^\ib -\ol\lambda^{\ib}_A\psi^A_{\mu}\label{supcovzb}
\end{align}
where the ordinary bosonic field strengths are denoted by:
\begin{align}
\cF^{ \Lambda}_{\mu\nu}= &  \partial_{[\mu}A^\Lambda_{\nu]}+\frac 12 f_{\Sigma\Gamma}{}^\Lambda A^\Sigma_{[\mu}\,A^\Gamma_{\nu]}+m^{\Lambda M}B_{M|\mu\nu}
\label{ flambda} \\
\cF_{M|\mu\nu}=& \partial_{[\mu}A_{M|\nu]}+\hat T_{\Lambda M}{}^N A^\Lambda_{[\mu}\,A_{N|\nu]} \label{fm} \\
 H_{M|\mu\nu\rho}=&  \partial_{[\mu}B_{M|\nu\rho]}+ T_{\Lambda M}{}^N A^\Lambda_{[\mu}\,B_{N|\nu\rho]} \label{hm}\\
\nabla_\mu z^i =& \partial_\mu z^i  + k^i_\Lambda A^\L_\mu  -  k^{iM}A_{M\mu}\,.
\end{align}

The mass matrices are given by:
\begin{align}
S_{AB}=& 0\\
%\frac \ii 2 (\sigma_x)_A{}^C \epsilon_{CB} \left(\xi^x_\Lambda L^\Lambda -\xi^{xM}L_M\right)\\
 W^{i[AB]}  = &\e^{AB} \left( k^i_\L \ol L^\L - k^{iM} \ol L_M \right) \\
 \cM_{jAkB}=&   \left[g_{\ib[j}\left( f^\Lambda_{k]}\,\ol k^\ib_\Lambda -  f_{M|k]}\,\ol k^{\ib M}\right)+\frac 1{2}g_{\ib j} \left(  L^\Lambda \nabla_k  \ol k^\ib_\Lambda -  L_M \nabla_k  \ol k^{\ib M}\right)\right]\epsilon_{AB}.
\end{align}
Note that the absence of a gravitino mass matrix $S_{AB}= 0$ is due to the absence of hypermultiplets and/or
Fayet-Iliopoulos terms in our setting.
%%%%%%%%%%%%%%%%%%%%%%%%%%%%%%%%%%%%%%%%%%%%%%%%%%%%%

\subsection{Supersymmetry transformation laws}
The supersymmetry transformation laws leaving the Lagrangian
invariant (up to total derivatives) are obtained from the superspace
curvatures obtained in our geometric approach.

For the fermion fields  we find:
\begin{align}
\delta \psi_{A\mu}=& \mathcal{D}_\mu\epsilon_A  +\epsilon_{AB}T^-_{\mu\nu}\gamma^\nu \epsilon^B +
\frac12 \epsilon_A Q^M \left(p_{Mi}\nabla_\mu z^i -\bar p_{M\ib}\nabla_\mu \bar z^\ib\right)+\nonumber\\
 &+ \left(A_A{}^B_\mu+\gamma_{\mu\nu}A'_A{}^{B\nu}\right)\epsilon_B+\nonumber\\
& -\frac12 Q^M  \left(p_{Mi}\ol \epsilon_B \lambda^{iB} -\bar p_{M\ib}\ol\epsilon^B \lambda^{\ib}_B\right)\psi_{A\mu}+\nonumber\\
&-\frac \ii 2 \left(Q_i \ol \lambda^{iB}\epsilon_B +Q_\ib   \ol \lambda^{\ib}_B\epsilon^B\right)\psi_{A\mu}\label{deltapsi}\\
\delta\lambda^{iA}=&\ii \tilde Z^i_\mu \gamma^\mu\epsilon^A + \cG^{-i}_{\mu\nu}\gamma^{\mu\nu}\epsilon_B\epsilon^{AB}+W^{iAB}\epsilon_B+\nonumber\\
&+\frac 1 2 \left(-C^i{}_{jk}\ol\lambda^{jA}\lambda^{kB}+\ii C^i{}_{\jb\kb}\ol\lambda^{\jb}_C\lambda^{\kb}_D\epsilon^{AC}\epsilon^{BD}\right)\epsilon_B+\nonumber\\
&+\frac 12 \lambda^{iA} Q^M \left(p_{Mi}\ol\lambda^{iB}\epsilon_B -\bar p_{M\ib}\ol\lambda^{\ib}_B\epsilon^B\right)\nonumber\\
&+\frac \ii 2 \left(Q_i \ol \lambda^{iB}\epsilon_B +Q_\ib   \ol \lambda^{\ib}_B\epsilon^B\right)\lambda_{iA}
-\Gamma^i{}_{jk}\ol\lambda^{kB}\epsilon_B\lambda^{jA}\label{deltalambda},
\end{align}
while for the boson fields we find:
\begin{align}
\delta V^a_\mu =& -\ii\ol\psi_{A\mu}\gamma^a\epsilon^A -\ii\ol\psi^A_{\mu}\gamma^a\epsilon_A\label{deltav}\\
\delta A^\Lambda_\mu =& -2L^\Lambda \ol\epsilon^A\psi^B_\mu \epsilon_{AB }-2\ol L^\Lambda \ol\epsilon_A\psi_{B\mu} \epsilon^{AB } +\nonumber\\
 &+\ii f^\Lambda_i \ol\epsilon^A\gamma_\mu \lambda^{iB}\epsilon_{AB}+ \ii \bar f^\Lambda_\ib \ol\epsilon_A\gamma_\mu \lambda^{\ib}_B\epsilon^{AB}\label{deltaalambda}\\
\delta A_{M\mu} =& -2L_M \ol\epsilon^A\psi^B_\mu \epsilon_{AB }-2\ol L_M \ol\epsilon_A\psi_{B\mu} \epsilon^{AB } +\nonumber\\
&+ \ii f_{Mi} \ol\epsilon^A\gamma_\mu \lambda^{iB}\epsilon_{AB}+ \ii \bar f_{M\ib} \ol\epsilon_A\gamma_\mu \lambda^{\ib}_B\epsilon^{AB}\label{deltaam}\\
\delta B_{M|\mu\nu}=& 2\ii P_M \left(\ol\psi_{A[\mu}\gamma_{\nu]} \epsilon^A +\ol\psi^A_{[\mu}\gamma_{\nu]}
 \epsilon_A\right)+p_{Mi}\ol\epsilon_A \gamma_{\mu\nu}\lambda^{iA}+\bar p_{M\ib}\ol\epsilon^A \gamma_{\mu\nu}\lambda^{\ib}_A
+\nn\\
&-(d_{\L\S M} A^\S_{[\mu} +\hat T_{\L M}{}^N A_{N[\mu})
\Bigl(\ii f^\Lambda_i \ol\epsilon^A\gamma_{\nu]}\lambda^{iB}\epsilon_{AB}+
\ii \bar f^\Lambda_\ib \ol\epsilon_A\gamma_{\nu]}\lambda^{\ib}_B\epsilon^{AB}+\nn\\
&-2L^\Lambda \ol\epsilon^A\psi^B_{\nu]} \epsilon_{AB }-2\ol L^\Lambda \ol\epsilon_A\psi_{B\nu]} \epsilon^{AB }\Bigr)
\label{deltab}\\
\delta z^i =&\ol\epsilon_A\lambda^{iA}\label{deltaz}
\end{align}
where:
\begin{align}
h_a =& \mez Q^M(p_{Mi} Z^i_a - \bar p_{M\ib} \ol Z^{\ib}_a )\nonumber\\
=& \frac\ii 4 Q^M \tilde H_M{}^{ bcd} \e_{abcd}
\\
T^-_{ab}=&   \left(\cN -\bar \cN\right)_{\L\S} L^\Lambda \left(\tilde\cF^{-\Sigma}_{ab}+\frac 18  \left( \nabla_i +  Q^M \,p_{Mi}  \right)f^\Sigma_j\ol\lambda^{iA}\gamma_{ab}\lambda^{jB}\epsilon_{AB}\right)\label{dpsi4}\\
\cG^{-i}_{ab}=& \frac\ii 2 g^{i\jb}(\cN -\ol \cN)_{\Lambda\Sigma} \bar f_\jb^\Lambda \left(\tilde\cF^{-\Sigma}+\frac 18 \left( \nabla_j +  Q^M \,p_{Mj}  \right)f^\Sigma_k\ol\lambda^{jA}\gamma_{ab}\lambda^{kB}\epsilon_{AB}\right)\\
A_{\mu|A}{}^B=& -\frac \ii 4 g_{i\jb}\left(\ol\lambda^\jb_A\gamma_\mu\lambda^{iB}-\delta_A^B \ol\lambda^\jb_C\gamma_\mu\lambda^{iC}\right)\\
A'_{\mu|A}{}^B=& \frac \ii 4\, g_{i\jb}\,\left(\ol\lambda^\jb_A\gamma_\mu\lambda^{iB}-\frac 12\delta_A^B \ol\lambda^\jb_C\gamma_\mu\lambda^{iC}\right)
\end{align}

%%%%%%%%%%%%%%%%%%%%%%%

\section{Some comments on the structure of the theory} \label{comments}
In this section we want to  make some observations on
the structure of the Lagrangian and its properties, taking into account the constraints found in superspace. A
complete list of the relations found on the $\sigma$-model and gauge structure is given in Appendix
\ref{relations}.

First of all we note that the kinetic Lagrangian \eq{lkin} does not contain an explicit propagation equation for
the tensors $B_{M|\mu\nu}$. Indeed, it is expressed in the first order formalism, through the auxiliary field
$Y_{M\mu}$, whose field equation gives
\begin{equation}
  Y_{M\mu}= \e_{\mu\nu\rho\sigma}\sqrt{-g} \tilde H_M{}^{\nu\rho\sigma}\label{tildeh0}\,.
\end{equation}

The dualization condition for the tensor field-strength in terms of the scalar 1-form degree of freedom
$Y_M^{(1)}$, \eq{dualization}, is obtained by the variation of the kinetic plus Pauli Lagrangian with respect to
${\tilde H}_{M\mu\nu\rho}$
$$\frac{\delta}{\delta {\tilde H}_{M\mu\nu\rho}}(\cL_k+\cL_{Pauli})=0\,,$$
giving the desired result
\begin{equation}
Y_{M\mu}= \e_{\mu\nu\rho\sigma}\sqrt{-g} \tilde H_M{}^{\nu\rho\sigma}
= -2\ii \left(p_{Mi}\tilde Z^i_\mu  -\bar p_{M\ib}\tilde{ \ol Z}^\ib_\mu \right)
\,.\label{dualization2}
\end{equation}
 Note however that the dualization is recovered in a simpler way if working in the geometric approach, since it is
 immediately
 found by solving the superspace Bianchi identity for the 3-form $H_M$ or, equivalently, by the superspace field equations
   from the
 superspace Lagrangian \eq{lkinrheo}, see eq.s \eq{starH}, \eq{tildeh}.

\par
 The equation of motion of the field $B_{M\mu\nu}$, taking into account the duality relation
 \eq{dualization2},
  gives:
  \begin{equation}
\tilde\cF^-_{M|ab} = \ol \cN_{M\Lambda}\tilde\cF^{-\Lambda}_{ab} = \ol \cN_{M X}\tilde\cF^{-X}_{ab}
+\ol \cN_{MN}\tilde\cF^{-N}_{ab}\label{fmlambda0}
\end{equation}
This relation, together with its complex conjugate for $\tilde\cF^+_{M|ab}$, allows to eliminate the Hodge-dual
field-strengths $F^M$ in terms of the fields $F_M$ contained in the tensor multiplets.
%Equivalently, the
%unphysical fields $F^M_{\mu\nu}$ and $Y_{M\mu}$ can be eliminated by expressing the theory in terms of the
%massive fields $A_M$ and $B_M$, after implementing the Higgs mechanism.
%
% see section \ref{emb}.

An important  observation is the following. As already touched on in Section \ref{susy4}, the
($2n_V+n_T$)-dimensional scalar sector of the off-shell theory is formulated in terms of a
($2n_V+2n_T$)-dimensional embedding manifold, $\cM_{(emb)}$, that can be endowed with a complex structure. We
note, as we will better see in the analysis of Appendix \ref{dualizationsection}, that the matrix $g_{i\jb}$ is
the metric of $\cM_{(emb)}$. Actually $g_{i\jb}$ does not satisfy the metric postulate. This is because of
\eq{torsionsigmamod}, found from the superspace constraints, that we rewrite here:
\begin{equation}
 \nabla_k g_{i\jb}= - g_{\ell
\jb}C^\ell{}_{ik} \,.
\end{equation}
However, if we modify the connection \begin{equation} \Gamma^i{}_{jk}\to
\mathring{\Gamma^i}{}_{jk}\equiv \Gamma^i{}_{jk}-C^i{}_{jk}
\end{equation}
then the new covariant derivative $\mathring{\nabla}[\mathring{\Gamma}]$  of
$g_{i\jb}$ is zero:
\begin{equation}
\mathring{\nabla}_k g_{i\jb}= \nabla_k g_{i\jb} + \, C^\ell{}_{ik} \,g_{\ell
\jb}=0\,.
\end{equation}
Furthermore, as observed in Appendix \ref{relations}, all the functions defined on $\cM_{(emb)}$ carry a weight
$(p, -p)$ with respect to the connections
\begin{equation}
\ii \cQ_i +\,Q^M p_{Mi}\,;\qquad \ii \cQ_\ib -\,Q^M \bar p_{M\ib}
\end{equation}
 respectively, that is with respect to the connection 1-form
 \begin{equation}
\ii \mathring{\cQ}\equiv \ii \cQ +\,Q^M (p_{Mi}\nabla z^i -\bar p_{M\ib}\nabla \ol z^\ib) = \ii \cQ +\frac \ii 2 Q^M Y_M^{(1)}
\end{equation}
 This allows to extend the definition of the $\mathring{\nabla}$ connection also to objects with non
 vanishing weight. Indeed if we define on $\cM_{(emb)}$ a total connection $\mathring{\Gamma}+i\mathring{\cQ}$ as
 follows
\begin{eqnarray}\label{redef}
    \mathring{\Gamma}^i_{jk}+i\,p\,\mathring{\cQ}_k\delta^i_j &\equiv &
     \Gamma^i{}_{jk}+\ii \,p\,\cQ_k\delta^i_j - C^i_{\;\; jk} + \,p\,Q^M p_{Mk}\delta^i_j\,,
  %    \mathring{\Gamma}^\ib_{jk} &\equiv &
%     \Gamma^\ib{}_{jk} -\ii C^\ib_{\;\; jk} =0\,,
 \end{eqnarray}
 then in terms of the covariant derivative $\cD$ defined in Appendix \ref{relations}, one has:
 \begin{equation}\label{redef2}
    \mathring{\nabla}_j\delta_k^i=
    \cD_j \delta_k^i - C^i_{\;\; jk}\,.
 \end{equation}
One can easily verify that all the constraints in Appendix \ref{relations} involving covariant derivatives of
$L^\Lambda, f^\Lambda_i$, when expressed in terms of $\mathring{\nabla}$ become exactly those defining
Special Geometry. Note that we also find, from the analysis of the superspace constraints:
\begin{eqnarray}
L_M&=& \cN_{M\L} L^\L\\
f_{Mi}&=&\ol\cN_{M\L} f^\L_i
\end{eqnarray}
which identifies them with the lower part of the symplectic vectors of Special Geometry $M_\L$ and $h_{\L i}$.

Now we recall that, as is well known, Special Geometry can be characterized  by the expression of its
K\"ahler--Hodge bundle whose curvature is (in the ungauged case)
\begin{equation}
\mathring{K} =   d \mathring{\cQ}=
 d\,\cQ -\ii d\left[\,Q^M (p_{Mi}d z^i -\bar p_{M\ib}d \ol z^\ib)\right]\label{kh}
\end{equation}
and by the
 curvature of the manifold, namely
\begin{equation}
\mathring{R}_{i\jb k\lb}(\mathring{\Gamma} + \ii \mathring{\cQ}) = g_{i\jb} g_{k\lb} + g_{k\jb}
g_{i\lb} - C_{ikm} C^m{}_{\jb\lb}\label{sg}\,,
\end{equation}
together with the relations:
\begin{eqnarray}
 \mathring{ \nabla}_{\lb} C^\kb{}_{ij}&=&0\label{SGC1}
   \\
 \mathring{ \nabla}_{[i} C^\kb{}_{\ell]j}&=&0\,.\label{SGC2}
\end{eqnarray}

 Eq. \eq{kh} implies the set of relations:
\begin{eqnarray}
K_{ij}&\equiv &  \nabla_{[i}\cQ_{j]}= \ii \nabla_{[i}(Q^M p_{M|j]})\label{k1}\\
K_{i\jb}&\equiv &  \nabla_i\cQ_\jb= \ii g_{i\jb}-\ii\left[ \nabla_{i}(Q^M \bar p_{M\jb }) +\nabla_{\jb}(Q^M p_{Mi })\right]\label{k2}
\end{eqnarray}
The simplest way to obtain  equation \eq{sg} is to perform the integrability of the equations
$\mathring{\nabla}_i f^\L_j = \ii \bar f^\L_\kb C^\kb{}_{ij}$ and $\mathring{\nabla}_\ib f^\L_j = g_{j\ib} L^\L$, together with their complex conjugates.
%Since our relations become coincident with the ones of special geometry when expressed in terms of
% $\mathring{\nabla}$, we then expect,

 If the  integrability conditions are implemented in equations (\ref{debareffe}) - (\ref{df4})
  using the $\Gamma + \ii \cQ$ connection
(corresponding to the covariant derivative $\nabla$), we find that   besides  $R^i{}_{j\kb l}$, also the Riemann
tensors $R^i{}_{jkl}$, $R^i{}_{j\kb\lb}$, $R^\ib{}_{jkl}$, $R^i{}_{jk\lb}$ $R^\ib{}_{j\kb l}$  are now different
from zero, in agreement with  the fact that $\cM_{(emb)}$ is not
 a K\"{a}hler manifold anymore.
 % They must be related to the corresponding ones of Special Geometry via a
% redefinition of the curvature, using the well-known recipe relating a curvature expressed
% in terms of a given connection to the one in terms of a different connection.
Indeed using \eq{redef} we have
\begin{eqnarray}
\mathring{R}{}^i{}_j(\mathring{\Gamma} + \ii \mathring{\cQ}) &=&
R^i{}_j(\Gamma+i\cQ) + \nabla (- C^i{}_j + \frac\ii 2 Q^M Y^{(1)}_M \d^i_j) + \nn\\
&&+(- C^i{}_k + \frac\ii 2 Q^M Y^{(1)}_M
\d^i_k)\wedge (- C^k{}_j + \frac\ii 2 Q^M Y^{(1)}_M \d^k_j)
%\\
%\mathring{R}{}^\ib{}_j(\mathring{\Gamma} + \ii \mathring{\cQ}) &=&
%R^\ib{}_j(\Gamma+i\cQ) + \nabla (\ii C^\ib{}_j )+\ii C^\ib{}_k
%\wedge \ii C^k{}_j
\label{rzero}
\end{eqnarray}
where $\G^i{}_j=\G^i{}_{jk}\nabla z^k$ and $C^i{}_j=C^i{}_{jk}\nabla z^k$ are (1,0)-forms while $Y^{(1)}_M
\equiv -2 \ii (p_{Mi} \nabla z^i-\bar p_{M\ib} \nabla \bar z^\ib)$ is the 1-form corresponding to the degrees of
freedom dual to the 3-form $H_M$. We found indeed, using \eq{kh}:
\begin{eqnarray}
R^k{}_{j|i\lb}&=&  -  2\delta^k_{(i}g_{j)\lb}+ C^k{}_{\lb\jb}C^\jb{}_{ij}- \nabla_\lb C^k{}_{ij}  \\
R^k{}_{j|i\ell}&=& \left(\nabla_{[i} C^k{}_{\ell]j} -  C^k{}_{m[i}C^m{}_{\ell]j}\right)= \buildrel       \circ\over{ \nabla}_{[i} C^k{}_{\ell]j} \\
R^k{}_{j|\ib\bar{\ell}}&=&0\\
R^\kb{}_{j|i\lb}&=& \ii\left(\nabla_{\lb} C^\kb{}_{ij} -  C^\kb{}_{\lb\jb}C^\jb{}_{ij}-2Q^M \ol              p_{M\lb}C^\kb{}_{ij}\right)\nn\\
&=&\ii \mathring{ \nabla}_{\lb} C^\kb{}_{ij}  \\
R^\kb{}_{j|i\ell}&=&  -\ii\left(\nabla_{[i} C^\kb{}_{\ell]j} + C^k{}_{j[i }C^\kb{}_{\ell ]k}+2Q^M        p_{M[i}C^\kb{}_{\ell ]j}\right) \nn\\
&=& - \ii \mathring{ \nabla}_{[i} C^\kb{}_{\ell]j} \\
R^\kb{}_{j|\ib\lb}&=&0
\end{eqnarray}
together with the constraints:
\begin{eqnarray}
g_{\kb[\ell }C^\kb{}_{i]j} &=&0\label{csymm}\\
\nabla_{[i}g_{j]\kb}&=&0
\end{eqnarray}
both implying, recalling the relations in Appendix \ref{relations}, that $C_{ijk}$ is completely symmetric. Note
that we can collect the Riemann tensors above in the 2-form expressions (where $C^i{}_j$, $C^\ib{}_j$ are $(1,0)$-forms):
\begin{eqnarray}
R^i{}_j&=&-2\delta^i_{(j} g_{k)\lb}\nabla z^k \wedge \nabla \ol z^\lb + \nabla C^i{}_j -C^i{}_k\wedge C^k{}_j-
C^i{}_\kb\wedge C^\kb{}_j\\
R^\ib{}_j&=& -\ii \nabla C^\ib{}_j +\ii C^\ib{}_k\wedge C^k{}_j+\ii
C^\ib{}_\kb\wedge C^\kb{}_j + Y_M^{(1)} \wedge C^\ib{}_j
\end{eqnarray}
One can check that expressing the curvatures $R^i{}_j$, $R^\ib{}_j$ of our embedding space in terms of
$\mathring{R}$ one finds that all the Riemann tensors are precisely related as in \eq{rzero} to the
corresponding quantities $\mathring{R}$. In particular,  $\mathring{R}{}^i{}_{\j\kb l}$
reduces to  eq. \eq{rzero}, as was to be expected. Requiring the vanishing of the other components one recovers
 the known relations of Special Geometry \eq{SGC1}, \eq{SGC2}.

A further observation is that in our theory we have, besides the quantities analogous to those of Special
Geometry, also the $\sigma$-model extra functions $P_M$, $p_{Mi}$, $Q^M$, $C^i{}_{jk}$, $k^{iM}$, corresponding
to extra structures and couplings arising from the presence of the vector-tensor multiplets. From this point of
view, we could think of our model as a particular gauging of $N=2$ supergravity coupled to vector multiplets,
where new structures and couplings have been introduced. In fact, if we let all these extra structures go to
zero, together with the electric and magnetic gauge coupling constants, all encoded in the symplectic quantities
$\mathcal{T}_\alpha$ defined in section \ref{emb}, we would recover the standard, ungauged special geometry.
This is in line with the approach of \cite{deWit:2005ub,de Wit:2007mt}, where the gauged theory coupled to
antisymmetric tensor fields was realized as a deformation of the ungauged theory coupled to vector multiplets,
in such a way that, setting the coupling constant to zero, the tensors get decoupled and the ungauged theory
retrieved. Such approach is somewhat complementary to the construction presented here, where our starting point
was the construction of an $N=2$ theory describing vector multiplets coupled to a number of vector-tensor ones.
Consistency of the theory then required the introduction of structures and extra fields which allowed us to
define a smooth limit to an ungauged theory where the antisymmetric tensors disappear in favor of their dual
scalar fields. In this limit $\cM_{(emb)}$ becomes the special K\"ahler scalar manifold, with metric
$g_{i\bar{\jmath}}$.
\par Therefore, although obtained
from a different perspective, the theory presented here can be
viewed as originating from a  deformation of an ungauged $N=2$
theory coupled to vector multiplets only.
 The global symmetry group
of this model is described by the isometry group ${\rm
Isom}(\cM_{(emb)})$ of the scalar manifold, endowed with a two-fold
action \cite{Gaillard:1981rj}: A non-linear action on the scalar fields
$z^i$ and a linear electric-magnetic duality action on the vector
field strengths and their magnetic duals, i.e. on
$\mathcal{F}^\alpha$.
 With reference to this ungauged theory, we could interpret the
symplectic matrices $(\mathcal{T}_{\alpha})_{\beta}{}^\gamma$ as the
generators of the gauge algebra embedded in the isometry algebra of
$\cM_{(emb)}$, namely expressed as linear combinations of the
generators $t_n$ of ${\rm Isom}(\cM_{(emb)})$ (with $n=1,\cdots , dim ({\rm Isom}(\cM_{(emb)}))$) through an
\emph{embedding tensor} $\theta_\alpha{}^n$:
\begin{eqnarray}
\mathcal{T}_{\alpha\beta}{}^\gamma\equiv
\theta_\alpha{}^n\,t_{n\beta}{}^\gamma\,,
\end{eqnarray}
where $t_{n\beta}{}^\gamma$ are the $2(1+n_V+n_T)\times
2(1+n_V+n_T)$ symplectic realization of the generators $t_n$ as
infinitesimal duality transformations on $\mathcal{F}^\alpha$.
Closure of the gauge group in ${\rm Isom}(\cM_{(emb)})$ is then
guaranteed by eq. (\ref{embalg}), which is a quadratic condition on
$\theta_\alpha{}^n$. Gauge invariance of the action and the absence
of anomalies further require the linear constraint (\ref{noanomaly})
on $\theta_\alpha{}^n$.\par A generic special K\"ahler manifold may
have no isometries at all. Somewhat implicit in our construction is
the presence in $\cM_{(emb)}$ of at least a number of
isometries $t_M$ which are parametrized by the scalars dual to the
antisymmetric tensor fields $B_{\mu\nu M}$.

As a final observation, we stress that all the relations obtained on the scalar sector are in fact relations on
the geometry of the embedding manifold $\cM_{(emb)}$ since they include all the coordinates, that is also the
auxiliary degrees of freedom dual to the tensors. Therefore, in order to obtain the geometry of the true
 $\sigma$-model underlying our theory we have to solve all these relations in terms of the physical scalar-field
 coordinates $(z^r,\bar z^\rb, P_M)$
 only. This requires a deep understanding of the embedding properties and will be the object of future
 investigation \cite{forth}.

%%%%%%%%%%%%%%%%%%%%%%%%%%%%%%%%%%%%%%%%%%%%%%%%%%%%%%%%%%%%%%%%%%%%%%%%

\appendix

%%%%%%%%%%%%%%%%%%%%%%%%%%%%%%%%%%%%%%%%%%%%%%%%%%%%%%%%%%%%%%%%%%%
%%%%%%%%%%%%%%%%%%%%%%%%%%%%%%%%%%%%%%%%%%%%%%%%%%%%%%%%%%%%%%%%%%%%

\section{Embedded algebra}\label{algebraemb}
We are going  to show that the $\cT_\alpha$, subect to \eq{bbi1} - \eq{bbi6} together with  \eq{anomaly1} -
\eq{anomaly4},
   close the algebra in the symplectic represntation:
 \begin{equation}
\left[\cT_{\alpha},\cT_\beta \right] = -\cT_{\alpha\beta}{}^\gamma \,\cT_\gamma\,.\label{embalg'}
\end{equation}
Eq. \eq{embalg'} corresponds to 3 matrix-relations:
\begin{enumerate}
\item[i)]
 \begin{equation}
\left[\cT_{\Lambda},\cT_\Sigma \right] = -\hat f_{\Lambda\Sigma}{}^\Gamma \,\cT_\Gamma\,.\label{embalg1}
\end{equation}
It requires the constraints:
 \begin{eqnarray}
\hat f_{\Lambda\Gamma}{}^\Delta \hat f_{\Sigma \Delta}{}^\Pi - \hat f_{\Sigma\Gamma}{}^\Delta \hat f_{\Lambda \Delta}{}^\Pi&=&-\hat f_{\Lambda\Sigma}{}^\Delta \hat f_{ \Delta\Gamma}{}^\Pi \label{embalg1a}\\
 f_{\Lambda X}{}^Y (\cT_{\Sigma})_{YZ} -f_{\Sigma X}{}^Y (\cT_{\Lambda})_{YZ}+ (X \leftrightarrow Z )&=&-\frac 12 f_{\Lambda\Sigma}{}^\Gamma (\cT_{\Gamma})_{XZ} \label{embalg1b}
\end{eqnarray}
Eq. \eq{embalg1a} is satisfied by the Free Differential Algebra, see \eq{hatbbi1}, while \eq{embalg1b} is
the condition for the tensor $(\cT_{\Gamma})_{XZ}$ to lie in the Chevalley-Eilenberg cohomology of the basis
of left-invariant 1-forms.

\item[ii)]
 \begin{equation}
\left[\cT_{\Lambda},\cT^P \right] = -\hat T_{\Lambda M}{}^P \,\cT^M\,,\label{embalg2}
\end{equation}
It corresponds to the constraints
 \begin{eqnarray}
\hat f_{\Lambda\Sigma}{}^ \Gamma \hat T_{\Gamma M}{}^P \,m^{\Delta M}  -\hat f_{\Lambda\Gamma}{}^\Delta \hat T_{\Sigma M}{}^P \,m^{\Gamma M}  &=&-\hat T_{\Lambda M}{}^P \hat T_{\Sigma N}{}^M \,m^{\Delta N} \label{embalg2a}\\
\hat  f_{\Lambda \Sigma}{}^\Gamma d_{(\Gamma \Delta) N}m^{NP}+ \hat  f_{\Lambda\Delta}{}^\Gamma d_{( \Sigma\Gamma) N}m^{NP}&=&\hat T_{\Lambda M}{}^P d_{( \Sigma\Delta) N}m^{NM}
\label{embalg2b}
\end{eqnarray}
Eq. \eq{embalg2a} is satisfied using \eq{bbi2}, \eq{bbi4} and the definitions of $\hat T$ and $\hat f$,
while eq. \eq{embalg2a} requires use of \eq{bbi6} and \eq{anomaly1}
\item[iii)]
 \begin{equation}
\left[\cT^{P},\cT^Q \right] = 0\,,\label{embalg3}
\end{equation}
It contains the 2 relations:
 \begin{eqnarray}
\hat T_{\Lambda N}{}^P m^{\Sigma N} \,\hat T_{\Sigma M}{}^Q m^{\Gamma M} -(P\leftrightarrow Q)&=&0  \label{embalg3a}\\
m^{MN}\, \hat T_{(\Lambda |N}{}^P\,\hat T_{(\Gamma ) | M}{}^Q -(P\leftrightarrow Q)&=&0\label{embalg3b}
\end{eqnarray}
Eq. \eq{embalg3a} is satisfied because of \eq{bbi4} and the definition of $\hat T$, implying the relation
$m^{\Sigma N} \,\hat T_{\Sigma M}{}^Q =0$; eq. \eq{embalg3b} requires $m^{MN}= m^{NM}$.
\end{enumerate}

\section{Superspace Bianchi identities and superspace parametrization of the curvatures}
\label{susyparam}

%%%%%%%%%%%%%%%%%%%%%%%%%%%%%%%%%%%%%%%%%%%%%%%%%%%%%%%%

The superspace Bianchi identities are:
\begin{align}
D \cR^a{}_b &= 0  \displaybreak[0] \\
D T^a &+ \cR^a{}_b V^b - \ii \ol \psi^A \g^a \rho_A - \ii \ol \psi_A
\g^a \rho^A = 0 \displaybreak[0]  \\
\na \rho_A &+ \qu \g_{ab} \cR^{ab} \psi_A - \imez K \psi_A = 0
\displaybreak[0] \\
\na \rho^A &+ \qu \g_{ab} \cR^{ab} \psi^A + \imez K \psi^A  = 0
\displaybreak[0] \\
\nabla F^\L &= \nabla L^\L \ol \psi^A \psi^B \e_{AB} + \nabla \ol L^\L \ol
\psi_A \psi_B \e^{AB} - 2 L^\L \ol \psi^A \rho^B \e_{AB} + \nn \\
& \quad - 2 \ol L^\L \ol \psi_A \rho_B \e^{AB} + m^{\L M} \left( H_M -
2\ii P_M \ol \psi_A \g_a \psi^A V^a \right)  \displaybreak[0] \\
\nabla F_M &= \nabla L_M \ol \psi^A \psi^B \e_{AB} + \nabla \ol L_M \ol
\psi_A \psi_B \e^{AB} - 2 L_M \ol \psi^A \rho^B \e_{AB} + \nn \\
& \quad - 2 \ol L_M \ol \psi_A \rho_B \e^{AB}  \displaybreak[0] \\
\nabla H_M &=2 \ii \nabla P_M \ol \psi_A \g_a \psi^A V^a - 2\ii P_M \left( \ol
\psi_A \g_a \rho^A + \ol \psi^A \g_a \rho_A \right) V^a + \nn \\
& \quad + 2P_M \ol \psi_A \g_a \psi^A \ol \psi_B \g^a \psi^B +  \nn \\
& \quad +\Bigl[d_{\L\S M}\left( F^\S - L^\S \ol \psi^A \psi^B \e_{AB} -
\ol L^\S \ol \psi_A \psi_B \e^{AB} \right) + \nn \\
& \,\qquad +\hat{T}_{\L M}{}^N \left( F_N - L_N
\ol \psi^A \psi^B \e_{AB} - \ol L_N \ol \psi_A \psi_B \e^{AB}
\right)\Bigr] \cdot \nn \\
& \quad \cdot \left( F^\L - L^\L \ol \psi^C \psi^D
\e_{CD} - \ol L^\L \ol \psi_C \psi_D \e^{CD} \right) \displaybreak[0] \\
D^2 z^i &= k^i_\L \left( F^\L - L^\L \ol \psi^A \psi^B \e_{AB} - \ol
L^\L \ol \psi_A \psi_B \e^{AB} \right)+ \nn \\
& \quad -   k^{iM} \left( F_M - L_M \ol \psi^A \psi^B \e_{AB} -\ol L_M
\ol \psi_A \psi_B \e^{AB} \right) \displaybreak[0]  \\
\na^2 \l^{iA} &= -\qu \g_{ab} \cR^{ab} \l^{iA} - \imez K \l^{iA} +
R^i{}_j \l^{jA} \displaybreak[0]\\
\na^2 \l^{\ib}_A &= -\qu \g_{ab} \cR^{ab} \l^{\ib}_A + \imez K
\l^{\ib}_A +  R^\ib{}_\jb \l^{\jb}_A \displaybreak[0]\
\end{align}
where:
\begin{align}
K&=\de\mathcal{Q}
\end{align}
is   the curvature associated to the potential $\cQ$, see eq.s \eq{k1}, \eq{k2}.

The superspace parametrizations of the curvatures are:
\begin{align}
\mathcal{T}^a &= 0 \label{par:tors} \displaybreak[0]\\
%%%%%%%%%%%%%%%%%
\cR_{ab} &=\tilde \cR_{abcd} V^c V^d - \ii \left(
\ol \psi_A \g_a \rho^A{}_{bc} + \ol \psi^A \g_a \rho_A{}_{bc} \right) V^c + \nn \\
& \quad - \ii   T^-_{ab}  \e_{AB} \ol \psi^A \psi^B
  - \ii   T^+_{ab}
\e^{AB} \ol \psi_A \psi_B + 2\e_{abcd}
A'_B{}^{A|c} \ol \psi_A \g^d \psi^B+ \nn \\
& \quad - S_{AB} \ol \psi^A \g_{ab} \psi^B - \ol
S^{AB} \ol \psi_A \g_{ab} \psi_B
\label{par:curv}\displaybreak[0]\\
%%%%%%%%%%%
\rho_A &= \rho_{A ab} V^a V^b +  \e_{AB} T^-_{ab} \g^b \psi^B V^a +
  h_a \psi_A V^a +
\ii S_{AB} \g_a \psi^B V^a+ \nonumber\\
& \quad +\frac 1 2 \psi_A Q^M \left( p_{Mi} \ol  \psi_B
\lambda^{iB} - \bar p_{M\ib}\, \ol  \psi^B \lambda^{\ib}_B\right) + \nonumber\\
&\quad +\left(A_{A\ a}^B +\g_{ab} A'^B_{A\
b}\right)\psi_B V^a
\label{par:rhol} \displaybreak[0]\\
%%%%%%%%%%%%%
\rho^A &= \rho^A_{ab} V^a V^b +  \e^{AB}
T^+_{ab} \g^b \psi_B V^a - h_a \psi^A V^a +
\ii \ol S^{AB} \g_a \psi_B V^a+\nonumber\\
&\qquad -\frac 1 2 \psi^A Q^M\left(p_{Mi} \ol  \psi_B
\lambda^{iB} -\bar p_{M \ib} \, \ol  \psi^B \lambda^{\ib}_B\right)  + \nonumber\\
&\qquad -\left( A_{B\ a}^A +\g_{ab}
{A'}^A_{B\ b}\right)\psi^B V^a
\label{par:rhor1}\displaybreak[0]\\
H_M &= \tilde H_{M|abc} V^a V^b V^c +  p_{Mi} \ol
\psi_A \g_{ab} \l^{iA} V^a V^b + \bar p_{M\ib} \ol \psi^A
\g_{ab} \l^{\ib}{}_A V^a
V^b  \label{par:HM}\displaybreak[0] \\
F^\L &= \tilde \cF^\L_{ab} V^a V^b + \ii f^\L_i
\ol \psi^A \g_a \l^{i B} \e_{AB} V^a + \ii \ol  f^\L_{\ib}
\ol \psi_A \g_a \l^{\ib}{}_B \e^{AB} V^a
\label{par:Flambda} \displaybreak[0]\\
F_M &= \tilde \cF_{M ab} V^a V^b + \ii f_{Mi}
\ol \psi^A \g_a \l^{i B} \e_{AB} V^a + \ii \ol  f_{M\ib} \ol
\psi_A \g_a \l^{\ib}{}_B
\e^{AB} V^a \label{par:FM}\displaybreak[0] \\
D z^i &= D_a z^i V^a + \ol \psi_A \l^{i A} \label{par:zi}\displaybreak[0] \\
D z^{\ib} &= D_a \ol z^{\ib} V^a + \ol \psi^A \l^{\ib}{}_A
\label{par:zibar}\displaybreak[0] \\
\na \l^{i A} &= \tilde \na_a \l^{i A} V^a + \ii
D_a z^i \g^a \psi^A + G^{i-}_{ab} \g^{ab} \e^{AB} \psi_B +
W^{i AB} \psi_B +\nn\\
& \quad +\frac 1 2 \l^{iA} Q^M \left(p_{Mj}
\ol  \psi_B \lambda^{jB} - p_{M\jb}\,
\ol  \psi^B \lambda^{\jb}_B\right)+ \nn\\
&\quad +\frac 1 2\sx[-C^i_{\ jk}\ol\l^{jA} \l^{kB} +\ii C^i_{\
\jb\kb}\ol\l^\jb_C \l^{\kb}_D \e^{AC}\e^{BD}\dx]\psi_B
\label{par:lambdal} \displaybreak[0]\\
\na \l^{\ib}{}_A &= \tilde \na_a \l^{\ib}{}_A V^a
+ \ii D_a \ol z^{\ib} \g^a \psi_A + G^{\ib +}_{ab} \g^{ab}
\e_{AB} \psi^B + \ol {W}^{\ib}{}_{AB} \psi^B +\nn\\
& \qquad -\frac 1 2 \l^\ib_A Q^M \,\left( p_{Mj}
\ol  \psi_B \lambda^{jB} - p_{M\jb}\, \ol  \psi^B \lambda^{\jb}_B\right)
+ \nn\\
&\quad -\frac 1 2\sx[ \ol C^{\ib}{}_{\jb\kb} \ol
\l^{\jb}_A \l^{\kb}_B +\ii \ol C^{\ib}{}_{jk} \ol
\l^{jC} \l^{kD} \e_{AC} \e_{BD} \dx]
\psi^B  \label{par:lambdar}\displaybreak[0]
\end{align}

%%%%%%%%%%%%%%%%%%%%%%%%%%%%%%%%%%%%%%%%%%%%%%%%%%%%%%%%
%%%%%%%%%%%%%%%%%%%%%%%%%%%%%%%%%%%%%%%%%%%%%%%%%%%%%%%%

\section{The superspace Lagrangian}\label{Lagrangian}

\begin{eqnarray}
\cL_{kin} &=& R^{ab} V^c V^d \e_{abcd}
- 4 \sx[ \ol \psi^A \g_a \rho_A - \ol \psi_A \g_a \rho^A \dx] V^a
+ 3\b_2 g_{i\jb} \ol\l^{iA} \g_b\l^\jb_A \,T_a \,V^aV^b+\nn\\
&& + \ii\sx[ \b_3 \left( \cN_{\L\S} \tilde\cF^{\L +}_{ab}
+ \ol \cN_{\L\S} \tilde\cF^{\L -}_{ab} \right)
\dx] \times \nn \\
&&\times \sx[ F^\S - \ii \left( f^\S_i \ol \l^{iA} \g_c \psi^B \e_{AB}
+ \bar f^\S_\ib \ol \l^{\ib}_A \g_c \psi_B \e^{AB} \right) V^c \dx] V^a V^b + \nn\\
&& - \frac{1}{24} \b_3 \left( \ol \cN_{\L\S} \tilde\cF^{\L -}_{ab} \tilde\cF^{\L - ab}
- \cN_{\L\S} \tilde\cF^{\L +}_{ab} \tilde\cF^{\L + ab} \right) \Omega + \nn \\
&& +  d_1 \cM^{MN} Y_{Na} \sx[ H_M - \left( p_{Mi} \ol \psi_A \g_{bc} \l^{iA}
+ \bar p_{M\ib} \ol \psi^A \g_{bc} \l^\ib_A \right) V^b V^c \dx] V^a + \nn \\
&&- \frac{1}{48} d_1 \cM^{MN} Y_{Ma} Y_N^a \Omega +\nn\\
&&+ \ii d_2 \cM^{MN}\sx[ H_M -\left(p_{Mi}\ol\psi_A \g_{ab}\l^{iA}
+ \bar p_{M\ib} \ol\psi^A\g_{ab} \l^\ib_A \right) V^a V^b \dx] \times \nn \\
&& \times \sx[ p_{Ni} \left(Dz^i - \ol \l^{iA} \psi_A \right)
- \bar p_{N\ib} \left( D \ol z^\ib - \ol \l^\ib_A \psi^A \right) \dx] + \nn \\
&& + \b_1 G_{i\jb} \sx[ \tilde Z^i_a \left( D \ol z^\jb - \ol \l^\jb_A \psi^A \right) + \tilde{\bar Z}^\jb_a
\left( D z^i - \ol \l^{iA} \psi_A \right) \dx] V^b V^c V^d \e_{abcd} \nn \\
&& + 2\g_1 \sx[ G_{ij}\tilde Z^i_a\left(D z^j - \ol\l^{jA}
\psi_A\right) + \ol G_{\ib\jb} \tilde{\bar Z}^\ib_a\left(D\ol z^\jb - \ol\l^\jb_A \psi^A\right)\dx] V^bV^cV^d\e_{abcd}\nn\\
&&-\frac 14 \sx[\b_1 G_{i\jb}\tilde Z^i_a \tilde{\bar Z}^{\jb ~a} +\g_1 \left(G_{ij}\tilde Z^i_a \tilde Z^{j~a}
+\ol  G_{\ib\jb}\tilde{\bar Z}^\ib_a \tilde{\bar Z}^{\jb ~a}\right)\dx]\Omega\nn\\
&& + \ii\b_2 g_{i\jb}\left(\ol\l^{iA}\g_a\nabla\l^\jb_A +
\ol\l^\jb_A \g_a\nabla\l^{iA}\right) V^bV^cV^d\e_{abcd} \label{lkinrheo}
\end{eqnarray}
Note that we have written in \eq{lkin} the kinetic terms  for the bosonic fields in first order formalism,
introducing the auxiliary fields $\tilde H_{Na}$, $\tilde\cF^{\Lambda}_{ab}$ $\tilde\cF_{M|ab}$, $\tilde Z^i_a$,
$\tilde{\bar Z}^{\jb }_a$ which turn out to be identified on shell with the corresponding supercovariant
field-strengths defined in \eq{supcovflambda} - \eq{supcovzb}.
\begin{eqnarray}
\cL_{Pauli} &=& -\ii\beta_3 F^\L \left(\cN_{\Lambda\Sigma}L^\Sigma\ol \psi^A \psi^B \e_{AB} +\ol\cN_{\Lambda\Sigma}\ol L^\Sigma\ol \psi_A \psi_B \e^{AB} \right)
+ \nn \\
&&-\beta_3F^\L\left(\ol\cN_{\Lambda\Sigma}f^\Sigma_i \ol\l^{iA} \g_a \psi^B \e_{AB}+ \cN_{\Lambda\Sigma}\bar f^\Sigma_\ib \ol\l^{\ib}_A \g_a \psi_B \e^{AB}\right) V^a+\nn\\
&&+ \b_7 F^\L \left(X_{\L ij} \ol\l^{iA} \g_{ab} \l^{jB} \e_{AB} - \ol X_{\L \ib\jb} \ol\l^{\ib}_A \g_{ab} \l^{\jb}_B \e^{AB} \right)V^aV^b +\nn\\
&&+\b_8 g_{i\jb}\left(D z^i \ol\l^\jb_A
\g^{ab}\psi^A + D\ol z^\jb \ol \l^{iA}\g^{ab}\psi_A\right)V^cV^d \e_{abcd}\nn\\
&&-\ii\b_9 Q^M \left(p_{Mi} D z^i - p_{M\ib} D\ol z^\ib\right) \ol\psi^A \g_a\psi_A V^a\nn\\
&& + \left( K_{i\jb k} D z^k - K_{i\jb\kb} D \ol z^{\kb} \right) \ol \l^{iA} \g^a \l^{\jb}_A \e_{abcd} V^b V^c V^d
\end{eqnarray}
\begin{eqnarray}
  \cL_{gauge}&=& \ii \d_1
 \left(S_{AB}\ol\psi^A\g_{ab}\psi^B +\ol
 S^{AB}\ol\psi_A\g_{ab}\psi_B\right) V^a V^b +\nn\\
  &&+\ii\d_2 g_{i\jb} \left(W^{iAB}\ol\l^\jb_A
 \g^a \psi_B + \ol W^\jb_{AB}\ol\l^{iA}\g^a \psi^B\right)V^bV^cV^d\e_{abcd}+\nn\\
&& + \d_6   \left(\cM_{iAjB}\ol\l^{iA}  \l^{jB} + \cM^{AB}_{\ib\jb}\ol\l^{\ib}_A\l^{\jb}_B\right)\Omega -\d_7 V \Omega
\end{eqnarray}
\begin{eqnarray}
\mathcal{L}_{CS} &=&
%a_\L^M B_M\left(\cF^\L +\frac
%12 m^{\L N}B_N\right) +
 a^{MN} \cF_M B_N +r_{\L\S}{}^M  dA_M  A^\L A^\S  +\nn\\
&&+ dA^\L\left( b_{\L\S\G} A^\S A^\G +
b_{\L\S}{}^M A^\S A_M + b_{\L}{}^{MN} A_M A_N\right) +\nn\\
&&+A^\L A^\S\left(t_{\L\S\G\D} A^\G A^\D +
t_{\L\S\G}{}^M A^\G A_M + t_{\L\S}{}^{MN}A_MA_N\right)
  \end{eqnarray}
 The expressions of the invariant tensors in $\cL_{CS}$ are fixed
 requiring gauge invariance of the full Lagrangian, as discussed below in section \ref{gaugeinv}. This is also a
 necessary condition for the Lagrangian to be supersymmetric.
\begin{align}
\cL_{4f} &= \frac{\ii}{2} \b_3 \left(L^\L \ol \psi^A \psi^B \e_{AB} + \ol L^\L \ol \psi_A \psi_B \e^{AB} \right) \left( \cN_{\L\S} L^\S \ol \psi^C \psi^D \e_{CD} + \ol \cN_{\L\S} \ol L^\S \ol \psi_C \psi_D \e^{CD} \right) + \nn \displaybreak[0] \\
& \hskip -15mm - \frac{\ii}{2} \b_3 \left( f^\L_i \ol \psi^A \g_a \l^{iB} \e_{AB} + \bar f^\L_{\ib} \ol \psi_A \g_a \l^\ib_B \e^{AB} \right) \left( \ol \cN_{\L\S} f^\S_j \ol \psi^C \g_b \l^{jD} \e_{CD} + \cN_{\L\S} \bar f^\S_\jb \ol \psi_C \g_b \l^\jb_D \e^{CD} \right) V^a V^b +  \displaybreak[0] \nn \\
& \hskip -15mm + \a_3 \left( f^\L_i \ol X_{\L\jb\kb} \ol \psi^A \g_c \l^{iB} \e_{AB} \ol \l^\jb_C \g_{ab} \l^\kb_D \e^{CD} - \bar f^\L_{\ib} X_{\L jk} \ol \psi_A \g_c \l^\ib_B \e^{AB} \ol \l^{jC} \g_{ab} \l^{kD} \e_{CD} \right) V^a V^b V^c + \displaybreak[0]  \nn \\
&\hskip -15mm + \a_4 g_{i\jb} \ol \l^{iA} \g_a \l^\jb_B \ol \psi_A \g_b \psi^B V^a V^b + \displaybreak[0] \nn \\
&\hskip -15mm + \a_5 \left( C_{ijk} \ol \l^{iA} \g_a \psi^B \ol \l^{jC} \l^{kD} \e_{AC} \e_{BD} -
 \ol C_{\ib\jb\kb} \ol \l^\ib_A \g_a \psi_B \ol \l^\jb_C \l^\kb_D \e^{AC} \e^{BD} \right) V^b V^c V^d \e_{abcd}
 + \displaybreak[0] \nn \\
&\hskip -15mm + \frac{1}{72}
\Bigl\{ \frac{3\ii}{16}(\cN -\ol\cN)_{\L\S} \left( C_{jkn}C_{i\ell m}g^ {m\mb}g^ {n\nb} \bar f^\L_\mb \bar f^\S_\nb\ol \l^{jA} \g_{ab} \l^{kB} \, \ol \l^{iC} \g^{ab} \l^{\ell D} \, \e_{AB} \e_{CD} + h.c.\right) + \displaybreak[0] \nn \\
&  \hskip -5mm  -\ii  \Bigl[ \left( \nabla_i  C_{jkl} + 2 Q^M p_{M i} C_{jkl} - 3 C^m{}_{ij} C_{k\ell m}  \right) \ol \l^{iA} \l^{jC} \ol \l^{kB} \l^{\ell D} \e_{AB} \e_{CD} + h.c.\Bigr] + \displaybreak[0] \nn \\
&  \hskip -5mm + 3\bigl[R_{i\jb k\lb}  - \frac{3}{2} g_{i\jb} g_{k\lb} +
 g_{i\lb} g_{k\jb}-\frac 12 g_{\ell\jb}\nabla_\lb C^\ell{}_{ik} -\frac 12 g_{k\kb}\nabla_i C^\kb{}_{\jb\lb}\bigr]\ol \l^{iA} \l^{kB} \ol \l^\jb_A \l^\lb_B \Bigr\} \, \Omega
\end{align}
where
$$\Omega \equiv V^aV^bV^cV^d \e_{abcd}$$
and \\
$\beta_1= \frac 23$, $\beta_2=- \frac 13$, $\beta_3= 4\ii$, $\beta_5=4$, $\beta_6 =-4$, $\beta_7= \frac
12$, $\b_8=1$, $\b_9= 4\ii$,
 $\gamma_1= \frac 13$, $d_1=-\frac 12$, $d_2=-1$,
$\d_1 =-4$, $\d_2= \frac 23$, $\d_6=\frac 16$, $\alpha_3 =\frac\ii 2$, $\alpha_4=2\ii$, $\alpha_5=-\frac 19$.

  %%%%%%%%%%%%%%%%%%%%%%%%%%%%%%
   %%%%%%%%%%%%%%%%%%%%%%%%%%%%%%
  %%%%%%%%%%%%%%%%%%%%%%%%%%%%%%

%%%%%%%%%%%%%%%%%%%%%%%
%%%%%%%%%%%%%%%%%%%%%%%

\section{Constraints on the $\sigma$-model and gauging}
\label{relations}
\begin{itemize}
\item
{\bf{ relations on the gauging:}}
\begin{align}
&k^{iM}=-2\ii m^{MN} \bar p_{N\jb} g^{i\jb}\\
& k^i_\Lambda m^{\Lambda M} =0\\
& k^i_\L L^\L  = k^{iM} L_M\\
%&\e_{AC} \left( f^{\L}_i W^{i (CB)} - 2 L^{\L}\ol S^{CB} \right) + \e^{BC}
%\left( \ol{f}^{\L}_{\ib} \ol W^{\ib}_{(AC)} - 2 \ol{L}^{\L} S_{AC} \right)
%= 0 \\
&f^\L_i \e_{AB} W^{iAB} + \bar f^\L_\ib \e^{AB} \ol W^\ib_{AB} + 4 m^{\L
  M} P_M = 0
\\
%& \epsilon_{AC} \left( f_{Mi} W^{i (CB)} - 2 L_M \ol S^{CB} \right)
%+ \epsilon^{BC} \left( \ol{f}_{M\ib} \ol W^{\ib}_{(AC)} - 2 \ol{L}_M S_{AC}
%\right) = 0 \\
& f_{Mi} \e_{AB} W^{iAB} + \bar f_{M\ib} \e^{AB} \ol W^\ib_{AB} = 0
\\
%&2P_M S^{AB} = p_{Mi} W^{i(AB)}\\
&Q^M p_{Mi} W^{i[AB]}  = Q^M \bar p_{M\ib} \ol  W^{\ib}_{[AB]} = 0\\
&g_{i(\jb}\left(\bar f^\Lambda_{\kb )}k^i_\Lambda -\bar f_{M|\kb )}k^{iM}\right)-g_{i\jb} C^i{}_{\kb\lb}\left( L^\Lambda\ol k^\lb_\Lambda - L_{M}\ol k^{\lb M}\right)=\nn\\
&=\frac 16 g_{i\jb}\left(\nabla_{\kb }k^i_\Lambda \ol L^\Lambda - \nabla_{\kb }k^{iM} \ol L_M\right)\\
&\nabla_{(i}\left(g_{j)\kb}k^\kb_\Lambda\right) = - C^k{}_{ij}g_{k\kb}\,k^\kb_\Lambda\\
&\nabla_{(i}\left(g_{j)\kb}k^{\kb M}\right) = - C^k{}_{ij}g_{k\kb}\,k^{\kb M}\end{align}
from which we deduce, using \eq{hmidef} and \eq{constrd}
\begin{equation}
\ol L^\Lambda\hat T_{\Lambda M}{}^NP_N Q^M =0
\end{equation}
The above relation is automatically satisfied if $Q^M \propto m^{MN}P_N$.
\begin{align}
&m^{MN}  P_M = -\frac \ii 2 Q^M \left( p_{Mi} k^{iN} - \bar p_{M\ib} k^{\ib N} \right) \\
&(\cN -\ol\cN)_{\L\S} f^\S_i (k^i_\L \ol L^\L -k^{iM} \ol L_M) = \nn \\
& \quad = \ii Q^M \left[p_{Mi} (k^i_\L + \cN_{\L N} k^{iN}) - \bar p_{M\ib}(k^\ib_\L + \cN_{\L N} k^{\ib N}) \right]
\end{align}

\item
{\bf{ relations on the gauge structure:}}
%\begin{equation}
%\cF^\L_{ab} = 2 \left( f^\L_i G^{i-}_{ab} + \bar f^\L_{\ib} G^{\ib +}_{ab} \right) + \ii \left( L^\L
%T_{ab}^+ + \ol L^\L T_{ab}^- \right)
%\end{equation}
\begin{equation}
3 \ii \tilde H_M{}^{abc} = \e^{abcd} \left(  p_{Mi}  Z^i_d - \bar p_{M \ib} \ol Z^\ib_d \right) \label{starH}
\end{equation}
\begin{equation}
  Y_{Ma}= \e_{abcd} \tilde H_M^{bcd}\label{tildeh}
\end{equation}
%
% \begin{eqnarray}
%\cF^\Lambda_{ab} &=& 2 \left( f^\Lambda_{i} G^{i-}_{ab} + \bar f^\Lambda_{\ib} G^{\ib +}_{ab}
%\right) + \ii \left( L^\Lambda T_{ab}^+ + \ol L^\Lambda T_{ab}^- \right)
%\\\cF_{Mab} &=& 2 \left( f_{Mi} G^{i-}_{ab} + \bar f_{M\ib} G^{\ib +}_{ab}
%\right) + \ii \left( L_M T_{ab}^+ + \ol L_M T_{ab}^- \right)
%\end{eqnarray}
%from which we get (substituting $T$ and $G$):
\begin{equation}
\tilde\cF^-_{M|ab} = \ol \cN_{\Lambda M}\tilde\cF^{-\Lambda}_{ab} \label{fmlambda}
\end{equation}
together with:
\begin{eqnarray}
\cN_{\L M}   L^\L &=&   L_M\\
\ol\cN_{\L M} f^\L_i &=& f_{Mi}\,.
\end{eqnarray}

\item
{\bf{ relations on the $\sigma$-model:}}

\begin{align}
\left(\cN -\bar \cN\right)_{\L\S}L^\Lambda \ol L^\Sigma &=-\ii\\
\left(\cN -\bar \cN\right)_{\L\S}f_i^\Lambda \bar f_\jb^\Sigma &=-\ii\,g_{i\jb}\\
\left(\cN -\bar \cN\right)_{\L\S}L^\Lambda   f_i^\Sigma &=0
\end{align}
\begin{equation}
\left(\cN - \ol \cN \right)^{-1|\L\S} = \ii \left( \bar L^\L L^\S + g^{i\jb}
f^\L_i \bar f^\S_\jb \right)
\end{equation}
We also find:
\begin{eqnarray}
\nabla_\jb \ol\cN_{\Lambda\Sigma}f^\Sigma_i &=& g_{i\jb}(\cN-\ol \cN)_{\Lambda\Sigma}L^\Sigma
\\
\nabla_{(i} \ol\cN_{\Lambda\Sigma}f^\Sigma_{j)} &=& \ii (\cN-\ol \cN)_{\Lambda\Sigma}\bar f^\Sigma_\kb C^\kb{}_{ij}
\end{eqnarray}
Furthermore:
\begin{align}
\left( \nabla_i +  Q^M \,p_{Mi}  \right) L^\Lambda &= f^\Lambda_i \\
 \left(\nabla_i - Q^M \,p_{Mi}  \right) \ol L^\Lambda &= 0\\
 \left( \nabla_{\ib} -  Q^M \bar p_{M\ib} \right) L^\Lambda &= 0 \\
 \left(\nabla_{\ib} +  Q^M \bar p_{M\ib} \right) \ol L^\Lambda
&= \bar f^\Lambda_{\ib}\\
\left( \nabla_i +  Q^M \,p_{Mi}  \right) L_M &= f_{Mi} \\
 \left(\nabla_i - Q^M \,p_{Mi}  \right) \ol L_M &= 0\\
 \left( \nabla_{\ib} -  Q^M \bar p_{M\ib} \right) L_M &= 0 \\
 \left(\nabla_{\ib} +  Q^M \bar p_{M\ib} \right) \ol L_M
&= \bar f_{M\ib}
\end{align}
\begin{align}
 \label{debareffe}\left( \nabla_{\jb} -  Q^M \bar  p_{M\ib} \right) f_i^\L &=   L^\L g_{i\jb}\\
\left( \nabla_{\jb} -  Q^M \bar p_{M\ib} \right) f_{Mi} &=   L_M g_{i\jb}
\end{align}
\begin{align}
\left( \nabla_{[i} +  Q^M \,p_{M[i}  \right) f_{j]}^\L &= 0 \label{df1}\\
\left( \nabla_{[i} +  Q^M \,p_{M[i}  \right) f_{M|j]} &= 0\label{df2}\\
\left( \nabla_i +  Q^M \,p_{Mi}  \right) f_j^\L &= - f^\L_{k} C^{k}{}_{ij} +\ii \bar f^\L_{\kb}
C^{\kb}{}_{ij} \label{df3}\\
 &= 8 f^\L_k G^k{}_{ij} + 8 \ii \ol L^\L T_{ij}\\
\left( \nabla_i +  Q^M \,p_{Mi}  \right) f_{Mj} &= - f_{Mk} C^{k}{}_{ij} +\ii \bar f_{M\kb}
C^{\kb}{}_{ij}\label{df4}\\
  &= 8 f_{Mk} G^k{}_{ij} + 8 \ii \ol L_M T_{ij}\\
  C^i{}_{[jk]} &= C^i{}_{[\jb\kb]} = 0
\end{align}
where
\begin{align}
T_{ij}=& \frac 18 (\cN -\ol \cN)_{\Lambda\Sigma}L^\Lambda \left( \nabla_i +  Q^M \,p_{Mi}  \right)f^\Sigma_j\\
\cG^k{}_{ij} =& \frac \ii 8 g^{k\kb}(\cN -\ol \cN)_{\Lambda\Sigma}\bar f_\kb^\Lambda \left( \nabla_i +  Q^M \,p_{Mi}  \right)f^\Sigma_j
\end{align}
In the relations above, the covariant derivative $\nabla$ is covariant with respect to the Levi-Civita
connection on the embedding manifold and to the K\"ahler connection $\cQ$, under which $L^\Lambda, L_M,
f^\Lambda_i, f_{Mi}$ have weight $+1$, their complex conjugates carrying weight $-1$. However, let us
observe that all the above relations can be written in terms of a new covariant derivative which includes
also a connection $Q^M (p_{Mi}\nabla z^i-\bar p_{M\ib}\nabla \ol z^\ib)$:
\begin{eqnarray}
\cD_i\equiv \nabla_i + p \,Q^M p_{Mi} \\
\cD_\ib\equiv \nabla_\ib - p \,Q^M  \bar p_{M\ib}
\end{eqnarray}
if we assume that the sections $L^\Lambda, L_M, f^\Lambda_i, f_{Mi}$ all carry the same weight +1 with
respect to the K\"ahler connection (their complex conjugates carrying the opposite weight).

\begin{align}
P_M &= - \frac 12\left[ 2 d_{(\L\S) M} L^\L \ol L^\S + \hat{T}_{\L M}{}^N \left(
\ol  L^\L L_N + L^\L \ol  L_N \right) \right]
\\
\nabla_i P_M &= p_{Mi} \nonumber\\
  &= - \frac 12 \left[ 2 d_{(\L\S) M} \ol L^\L f_i^\S + \hat{T}_{\L
  M}{}^N \left( \ol L^\L f_{iN} + f_i^\L \ol L_N \right) \right]\label{hmidef}
\end{align}
together with the constraints
\begin{align}
& d_{(\L\S) M} L^\L L^\S + \hat{T}_{\L M}{}^N L^\L L_N = 0\,,\\
& 2 d_{(\L\S) M} L^\L f_i^\S + \hat{T}_{\L M}{}^N \left( L^\L f_{iN} +
f_i^\L L_N \right) = 0\label{constrd}
\end{align}
\begin{align}
 \nabla_{[i} p_{Mj]}& = 0 \\
 \nabla_i p_{Mj}  &= - p_{Mk} C^{k}{}_{ij} \\
 \nabla_i p_{Mj} &= 8 p_{Mk} G^k{}_{ij}
\\
 P_M \ol T_{\ib\jb} &=  - \frac{1}{8}p_{Mk} C^k{}_{\ib\jb}\\
 &=- \frac{\ii}{8} \sx[ d_{(\L\S)M} \bar f^\L_\ib \bar f^\S_\jb + \hat{T}_{\L M}{}^N \bar f^\L_{(\ib} \bar f_{N|\jb)} \dx]
\end{align}
\begin{align}
\nabla_i \bar p_{M\jb} &= \nabla_\jb p_{Mi}\\
\nabla_i \nabla_\jb P_M &= \frac 12\sx[ 2d_{(\L\S)M} f^\L_i \bar f^\S_\jb + \hat{T}_{\L
  M}{}^N \left( f^\L_i \bar f_{M\jb} + \bar f^\L_\jb f_{Ni} \right) \dx] \nn\\
&= \frac 12   g_{i\jb} P_M
\end{align}

\begin{align}
   \cM^{MN} p_{Mi} p_{Nj} =&  2 G_{ij}   \\
   \cM^{MN} p_{Mi} \bar
 p_{N\jb} =& -2 G_{i\jb}+2 g_{i\jb}
\end{align}
\begin{eqnarray}
&&-\frac 13\left(  \na_k g_{i {\jb}} -  Q^M p_{Mk} g_{i\jb}\right) =\ii K_{i\jb k}\\
%&&-\frac 19 \ol C_{\jb i k} = \frac \ii 2\left(\alpha_6 + \alpha_7\right) J_{i\jb k} \\
%&&\frac 14 \left(\alpha_6 -3 \alpha_7\right) J_{j\kb i}=\left(\a_3 - \ii \b_7\right)\bar f^\L_\kb X_{\L ij}
% +\nn\\
%&&+ \left(\tilde\a_3
%-\ii \g_7\right)\bar f_{M \kb}X^M_{ij}
%\\
%&&-\frac 12 Q^M g_{i[\jb}p_{M|\kb ]}\left( k^i_\Lambda \ol L^\Lambda -  k^{iM} \ol L_M\right)\\
%
%
%g_{i(\jb}\nabla^+_{\kb )}W^i&=& C_{\jb\kb\lb}\bar W^\lb+\nn\\
%&-&\frac 23\,g_{i(\jb}
%\left[\left(\nabla^+_{\kb)} k^i_\Lambda\right)L^\Lambda - \left(\nabla^+_{\kb)} k^{iM}\right)L_M\right]\\
&&\nabla_k g_{i\jb}= - g_{\ell \jb}C^\ell{}_{ik}\label{torsionsigmamod}\\
&&X_{\Lambda ij}
%+\IP_{\Lambda M}X^M{}_{ij}
=\ii (\cN -\bar \cN)_{\Lambda\Sigma}\bar f^\Sigma_\lb \bar g^{k\lb}g_{i\kb}C^\kb{}_{jk}\end{eqnarray}
\begin{eqnarray}
  g_{i \kb} C^{\kb}{}_{(jk)} &=&   \frac 13\left(C_{ijk} +C_{jki}+C_{kij}\right)\\
  &=&   f^\L_{k} X_{\L (ij)}
  \\
C_{ijk} &=& C_{kji}
\\
 8\ii  g_{\ell\ib}\mathcal{G}^\ell{}_{jk} + g_{\ell\ib}C^\ell{}_{jk} +\bar f^\L_\ib  X_{\L (jk)} &=& 0\\
 g_{k\kb}\nabla_\lb C^k{}_{ij}&=&g_{j\jb}\nabla_i C^\jb{}_{\kb\lb}
 %+ \gamma_7 f_{M i} \ol X^M{}_{(\jb\kb)} \\
 %\ii \b_2 g_{i \lb} \ol C^{\lb}{}_{(\jb\kb)} &=&  \a_5 C_{i\jb\kb}
\end{eqnarray}

From the above relations we find
\begin{eqnarray}
&&\hat{T}_{\L M}{}^N (\cN - \ol \cN)_{N \S}   L^\L   f^\S_k C^k{}_{\ib\jb} = 0\\
&& g_{k\jb}C^\jb{}_{j\ell} g^{\ell \ib}= C^\ib{}_{jk}
\end{eqnarray}

\end{itemize}

\subsection{Constraints from gauge invariance}
\label{gaugeinv}

The Lagrangian must be gauge-invariant up to total derivatives. In checking this property, in particular two
different sectors in the gauge variation of the Lagrangian do not depend on scalar fields and hence should
cancel out each other: the first comes from $\cL_{CS}$ and the second from the topologial sector of $\cL_{Kin}$:
$$\cL_{Kin,top} \propto \re_{\L\S} F^\L F^\S $$ which contributes non trivially when we consider gaugings with a non trivial action on the kinetic matrix of the gauge fields, that is gaugings with a non block-diagonal symplectic embedding.
To have gauge invariance of $\cL_{CS}+\cL_{kin}$, we must take into account the possible constant contributions
from the gauge variation of the kinetic Lagrangian \cite{de Wit:1984px}.

The gauge transformations of the fields under vector-gauge transformations with parameter $\e^\Lambda$, $\e_M$,
together with the symplectic embedding and the gauge transformation of $\cN$, are given in Section \ref{emb}.

We found the following set of conditions on the constant couplings:
\begin{eqnarray}
%(\cT_{\Gamma})_{\L\S}&=& \cT_{X(YZ)}\delta_\Gamma^X \delta_{(\Lambda\Sigma)}^{YZ}\\
%(\cT^M)_{\L\S}&=& 2 m^{MN}d_{(\Lambda\Sigma)N}\\
%a_\Lambda^M&=&0\\
a^{MN}&=& 4 m^{MN}\\
b_{\L[\S\G]}&=& -\frac 23 \left[ (\cT_{\Gamma})_{\L\S}-(\cT_{\Sigma})_{\L\Gamma}\right]
\\
b_{(\L\S)}{}^M&=&-4m^{MN}d_{(\Lambda\Sigma)N}\\
b_\Lambda{}^{[MN]}&=& 2\, \hat T_{\Lambda P}{}^{N}m^{MP}\\
r_{[\L\S]}{}^M&=& 2\, m^{MN}d_{[\Lambda\Sigma]N}\\
t_{[\L\S\G\D]}&=& \frac 12 f_{[\Lambda\Sigma}{}^\Omega(\cT_{\Gamma})_{\Delta]\Omega}  \\
t_{[\L\S\G]}{}^M&=&-2 m^{MN} d_{\Delta [\Lambda |N}f_{\Sigma \Gamma]}{}^\Delta\\
t_{[\L\S]}{}^{[MN]}&=& \hat T_{[\Gamma | P}{}^{N}m^{MP}
 f_{\L\S}{}^\G
\end{eqnarray}

%%%%%%%%%%%%%%%%%%%%%%%

\section{The vector-tensor $\sigma$-model metric}
\label{dualizationsection} Let us start from the (ungauged) kinetic term of special geometry:
\begin{eqnarray}
\cL_{kin} &=& g_{i\jb} \partial_\mu z^i \partial^\mu \ol z^\jb \label{lagrsg}
\end{eqnarray}
In light of eq. \eq{starH}, we want to dualize
\begin{equation}
- 2 \ii \left( p_{Mi} dz^i - \bar p_{M\ib} d \ol z^\ib \right) \equiv Y^{(1)}_M
\end{equation}
To this aim, let us assume the following Ansatz for the metric:
\begin{equation}
g_{i\jb}= \hat G_{i\jb} + \cM^{MN} p_{Mi} \bar p_{N\jb}
\end{equation}
and let us decompose $p_{Mi} dz^i$ into real and imaginary parts:
\begin{align}
p_{Mi}d z^i &= \frac{1}{2} \left(p_{Mi} d z^i + \bar p_{M\ib } d \ol z^\ib \right) + \frac{\ii}{4} Y^{(1)}_M \label{hdz} \nn \\
&= \frac{1}{2} \nabla P_M + \frac{\ii}{4} Y^{(1)}_M
\end{align}
In terms of the new variables, the Lagrangian \eq{lagrsg} reads:
\begin{equation}
\cL_{kin} = \frac{1}{2} G_{ij} \partial_\mu z^i \partial^\mu z^j +
 G_{i\jb} \partial_\mu z^i \partial^\mu \ol z^\jb + \frac{1}{2} G_{\ib\jb}
 \partial_\mu \ol z^\ib \partial^\mu \ol z^\jb + \frac{1}{16} \cM^{MN} Y_{M \mu} Y_N{}^\mu \label{lagrdual}
\end{equation}
where
\begin{eqnarray}
G_{ij}&=& \frac 12 \cM^{MN} p_{Mi} p_{Nj}\\
G_{\ib\jb}&=& \left(G_{ij}\right)^*\\
G_{i\jb}&=& g_{i\jb} -\frac 12\cM^{MN} p_{Mi} \bar p_{N\jb}
\end{eqnarray}
To perform the dualization on the vector-tensor multiplet sector, we introduce the Lagrange multiplier
$Y_{M\mu}$ and add to the Lagrangian \eq{lagrdual} the term
\begin{equation}
- \frac{1}{8} \cM^{MN} Y_{M\mu} H_{M \nu\rho\sigma} \epsilon^{\mu\nu\rho\sigma}
\end{equation}
Varying then the new Lagrangian with respect to $Y_{M\mu}$ we obtain
\begin{equation}
Y_M{}^\mu =  H_{M \nu\rho\sigma} \epsilon^{\mu\nu\rho\sigma}
\end{equation}
We observe that we are using a redundant parametrization of the scalar manifold in terms of the $2n_V + 2n_T$
coordinates $z^i, \overline z^\ib$. Actually the scalar manifold has real dimension $2n_V + n_T$ and we want
parametrize it with general coordinates $\phi_\a$. In order to know the constraints on the geometry coming from
supersymmetry, it is then necessary to pull back all the differentials appearing in the relations found from the
Bianchi Identities and the Lagrangian in terms of the differentials $\de \phi^\a$. The simplest way to perform
the pull-back is to choose a particular system of coordinates: $\phi^\a = \{ z^r, \ol z^{\bar r}, P_M \}$ so
that
\begin{align}
\der_\a z^i &= \{ \d^i_r, 0, \der^M z^i \equiv \xi^{Mi} \} \\
\der_\a \ol z^\ib &= \{ 0, \d^{\bar r}_\a, \der^M \ol z^\ib \equiv \ol \xi^{M\ib} \} \\
\der_\a P_M &=   p_{Mi} \der_\a z^i +   \bar p_{M\ib} \der_\a \ol z^\ib = \{  p_{Mr},  \bar p_{M\bar r}, \d_{MN} \}
\end{align}

\begin{equation}
G_{\a\b} =
\begin{pmatrix}
g_{\a\b} + \mez \cM^{MN} p_{M\a} p_{N\b} & \cM^{MN} p_{N\b} \\
\cM^{NR} p_{R\a} & \cM^{MN}
\end{pmatrix}
\end{equation}
A detailed analysis of the $\sigma$-model geometry will be presented in \cite{forth}.

\end{document}